\documentclass[english,secnumarabic,amssymb, nobibnotes, aps, pre, preprint]{revtex4-1}
\usepackage[english]{babel}
\usepackage[T1]{fontenc}
\usepackage{amsmath}
\usepackage{amssymb}
\usepackage{graphicx}
\usepackage{esint}
\usepackage{tikz}
\usetikzlibrary{chains}
\usetikzlibrary{arrows}
\usepackage[caption=false]{subfig}
\usepackage{float}

\makeatletter

\@ifundefined{textcolor}{}
{%
 \definecolor{BLACK}{gray}{0}
 \definecolor{WHITE}{gray}{1}
 \definecolor{RED}{rgb}{1,0,0}
 \definecolor{GREEN}{rgb}{0,1,0}
 \definecolor{BLUE}{rgb}{0,0,1}
 \definecolor{CYAN}{cmyk}{1,0,0,0}
 \definecolor{MAGENTA}{cmyk}{0,1,0,0}
 \definecolor{YELLOW}{cmyk}{0,0,1,0}
}

\newcommand*{\rom}[1]{\expandafter\@slowromancap\romannumeral #1@}

\makeatother

\usepackage{letltxmacro}

\LetLtxMacro{\ORIGselectlanguage}{\selectlanguage}
\makeatletter
\DeclareRobustCommand{\selectlanguage}[1]{%
  \@ifundefined{alias@\string#1}
    {\ORIGselectlanguage{#1}}
    {\begingroup\edef\x{\endgroup
       \noexpand\ORIGselectlanguage{\@nameuse{alias@#1}}}\x}%
}
\newcommand{\definelanguagealias}[2]{%
  \@namedef{alias@#1}{#2}%
}
\makeatother

\usepackage[english]{babel}
\definelanguagealias{en}{english}
\begin{document}

\title{Analysis of Time Scale to Consensus in Voting Dynamics with more than Two Options}

\author{Degang Wu}
\author{Kwok Yip Szeto}
\affiliation{Department of Phyiscs, The Hong Kong University of Science and Technology, Clear Water Bay, Hong Kong, HKSAR, China}
\begin{abstract}
We generalize a binary majority-vote model on adaptive networks to its plurality-vote counterpart and analyze the time scale to consensus when voters are given more than two options.
When opinions are uniformly distributed in the population of voters in the initial state, we find that the time scale to consensus is shorter than the binary vote model from both numerical simulations and mathematical analysis using the master equation for the three-state plurality-vote model. 
When intervention such as opinion conversion is allowed, as in the case of sudden change of mind of voter for any reason,  the effort needed to push the fragmented three-opinion population in the thermodynamic limit to the consensus state, measured in minimal intervention cost, is less than that needed to push a polarized two-opinion population to the consensus state, when the  degree ($p$) of homophily is less than 0.8.
For finite system, the fragmented three-opinion population will spontaneously reach the consensus state, with faster time to consensus, compared to polarized two-opinion population, for a broad range of $p$.
\end{abstract}
\maketitle

\section{Introduction}

Interest in problems of voting dynamics and opinion formation is not limited to social-political studies, as many models constructed by physicists and mathematicians have been designed to estimate the time needed to reach consensus. 
Examples are the voter model~\cite{clifford_model_1973, holley_ergodic_1975}, majority-rule model~\cite{galam_minority_2002}, Sznaj model~\cite{katarzyna_opinion_2000,sabatelli_non-monotonic_2004}, Axelrod's model~\cite{axelrod_dissemination_1997,castellano_statistical_2009} etc. 
For a review of major models refer to \cite{sen_sociophysics:_2013}.
One of the key question concerns the evolution of opinion in a multi-agent system, where the agents can be modeled by ``particles'' with special attributes and interactions that can also be changing with time. 
The agents, voters, or particles are modeled as nodes in a social network, with links between nodes specifying their interaction. 
Since a changed opinion (or attribute) of the agent can induce changes to the connections with neighboring nodes, while a changed connection can also induce a change to the opinion of the agent, the entire system of interacting agents is therefore a co-evolving network with both nodes and links changing. 
The goal of opinion formation is to count the number of agents holding a particular opinion as a function of time, but the fact that the links connecting nodes are also changing with time implies that we are addressing a problem of great complexity defined on a ``social network'' of evolving topology. 
The complexity of this problem is further accentuated by the deadline imposed on the specific election.  
Consequently, the usual studies of time scale to reach consensus in voting models must be rephrased in terms of the speed to consensus. 
For example, a party in an election may win in the long run, but in the short run, such as at the deadline for counting the vote, another party may have more votes and end up winning. 
Therefore a comparison of time scales for the opinion formation process is very important in application. 
In this paper, we address this question of time scales from the perspective of three-state plurality-vote model.
Does having more options to voters mean harder to reach consensus? 
The answer we obtained from both numerical simulation and the mathematical analysis of the master equation shows an intriguing result: voters with options in a three party race will reach consensus faster than when there are only two parties.  

While agent-based simulations were frequently employed to study co-evolving opinion dynamics, the extension to large scale usually encounters problems due to the complexity of the model with updating rules that are complex if the model is realistic~\cite{demirel_moment-closure_2014}. 
Therefore a complementary approach is to build simpler, but mathematically amenable models such as the one proposed by Benczik et al.~\cite{benczik_opinion_2009}, so as to extract valuable insights in the qualitative behaviors observed in simulation. 
The opinion in these mathematical models can either be a discrete~\cite{galam_minority_2002,katarzyna_opinion_2000,de_oliveira_isotropic_1992}, or continuous variable \cite{hegselmann_opinion_2002}, while the exact interpretation of the opinion is very flexible, depending on the context in application. 
For example, opinions could be political views to adhere to, sports teams to support, musical styles people enjoy, and so on. For the discrete models, most research focuses on the simplest two-state model, i.e., the opinion is a ``yes'' or ``no'' response to an issue. 
Recent work suggests that the time to consensus increases with the number of available opinions~\cite{wang_freezing_2014}, while other numerical work~\cite{wu_three-state_2014} suggests otherwise. 
Since the nature of the increase or decrease in time to consensus is still unclear, we like to clarify this issue for the case of three-opinion model. 
The conclusion of our investigation must also be tested for large population, so that scaling behavior of the time to consensus can be addressed.

Our starting point is to generalize the binary majority-vote model on adaptive networks~\cite{benczik_opinion_2009} to plurality-vote model with more than two states. 
Our approach is mainly numerical, but we also analyze our numerical results to achieve a better understanding of the mechanism behind the various time scales to consensus.
Our analytical work confirms the scaling relation with the population size $N$ with our numerical results.  
Our work is complementary to those in Refs.~\cite{gekle_opinion_2005, galam_drastic_2013}, which investigated three opinion system with discussion-group-dynamics.
The focus was on the dominance of minority opinion due to hidden preferences in case of a tie in voting and the size of discussion group is fixed so as to allow full analytical treatment.

\section{Model}

Our model consists of $N$ agents (nodes). Each agent $i$ carries an opinion $\sigma_{i}=1,2$ or $3$, with $i=1,2,\dots,N$. 
In each time step, we randomly choose an agent $i$ to be updated. 
Temporary links will be formed between $i$ and other agents in the population, according to a probability $p$ and $q$, which are constants for the whole population.
We go through all possible edges between $i$ and $j$, where $j=1,2,\dots N$ and $j\neq i$. 
If $\sigma_{i}=\sigma_{j}$, then a link will be formed between the two nodes with a probability $p$, called the \textit{degree of homophily}.
If $\sigma_{i}\neq\sigma_{j}$, a link will be formed with a probability $q\equiv1-p$.
Once we have decided all the temporary links between agent $i$ and all other agents, we update $i$ using the following rule: we count the number of the three opinions in $i$'s temporary neighborhood.
If there is a plurality opinion in the temporary neighbors $(v)$, then we update the agent $i$'s opinion by $\sigma_{i}=v$; otherwise $\sigma_{i}$ remains unchanged.
Here, by \emph{plurality}, we mean the situation when the number of one opinion is larger than the number of any of the other opinions.
Therefore, in this work, majority is a special case of plurality.
This update rule is very similar to the majority rule model~\cite{galam_minority_2002}.
After the update, all temporary links are eliminated. 
The temporary nature of the link formation process renders our model amenable to analytical treatment. 
The structure of our model is similar to the two-opinion model of Benczik et al.~\cite{benczik_opinion_2009}, so that the temporary nature of the link formation renders our model amenable to mean field analysis. 
In our model, large $p$ could indicate that individuals are more likely to hear from people holding the same opinion (homophilic) or support the same political party. 
Small $p$ may represent the situation where individuals are more likely to interact with people with different and diverse background (heterophilic) or not satisfied with the original opinion or party, and are seeking for a different opinion. 

First we study the system in the thermodynamic limit, using analytical approaches such as analysis of attractors and basins of attraction of the master equation.
Then we investigate the finite size effects by numerically integrating the master equation for finite population and performing Monte Carlo simulations. 
We like to know, in the thermodynamic limit, if a population with certain value of $p$ and at a particular \textit{fragmented state} (three or more opinions co-exist) will converge to the \textit{consensus state} (there is only one opinion in the population) and what is the time scale of convergence.
We also like to know, if intervention is allowed, what is the minimal effort needed to push a fragmented population to the consensus state. 
Finally, we like to know the impact of the finite size effects on the system.
In finite size systems, a population will always converge to the consensus state, but the time it takes, called the \textit{time to consensus}, could vary wildly.
The distribution of time to consensus could have significant implications in the behaviors of the system being modeled. 
For real election, which has a deadline for voting, the convergence time is of great practical importance, as they will determine which party will win the election. 

\section{Thermodynamic Limit}
First we denote by $N_1$ ($N_2$, $N_3$) the number of agents holding opinion 1 (2, 3). 
Since $N_1+N_2+N_3=N$, $N_1$ and $N_2$ suffice to describe the state of the system in the configuration space. 
Therefore we denote by $P(N_1,N_2)$ the (time-dependent) probability that the system is in state $(N_1,N_2)$.
From the definition, one can derive the master equation:
\begin{equation}
\begin{aligned}
 &\partial_{t}P(N_1,N_2)=-\sum_{i\neq j,i,j=1}^3 W_{i\rightarrow j}(N_i,N_j)P(N_i,N_j)\\
 +&\sum_{i\neq j,i,j=1}^3 W_{i\rightarrow j}(N_i+1,N_j-1)P(N_i+1,N_j-1),
\end{aligned}
\label{eq:master_eq_short}
\end{equation}
where 
\begin{equation}
\begin{aligned}
&W_{i\rightarrow j}(N_i,N_j)= \frac{N_i}{N}\sum_{l=0}^{N_i-1}\sum_{l'=0}^{N_j}\sum_{l''=0}^{N_k}&\left[\Theta(l'-l)\Theta(l'-l'')\right.\\
&\left.B_{N_{i-1},p}(l)B_{N_j,q}(l')B_{N_k,q}(l'')\right],
\end{aligned}
\label{eq:Wi2j}
\end{equation}
where $k\neq i\neq j$ and 
\begin{equation}
\begin{aligned}
B_{n,p}(l)={n \choose l}p^{l}(1-p)^{n-l},
\label{eq:BNp}
\end{aligned}
\end{equation}
and $\Theta(\cdot)$ is the unit step function.
\begin{figure}

	\subfloat[$p=0.35$\label{fig:trans_prob_regions_p_0.35}]{%
	\includegraphics[trim=0.35in 0.41in 0.43in 0.3in,clip,width=0.5\columnwidth]{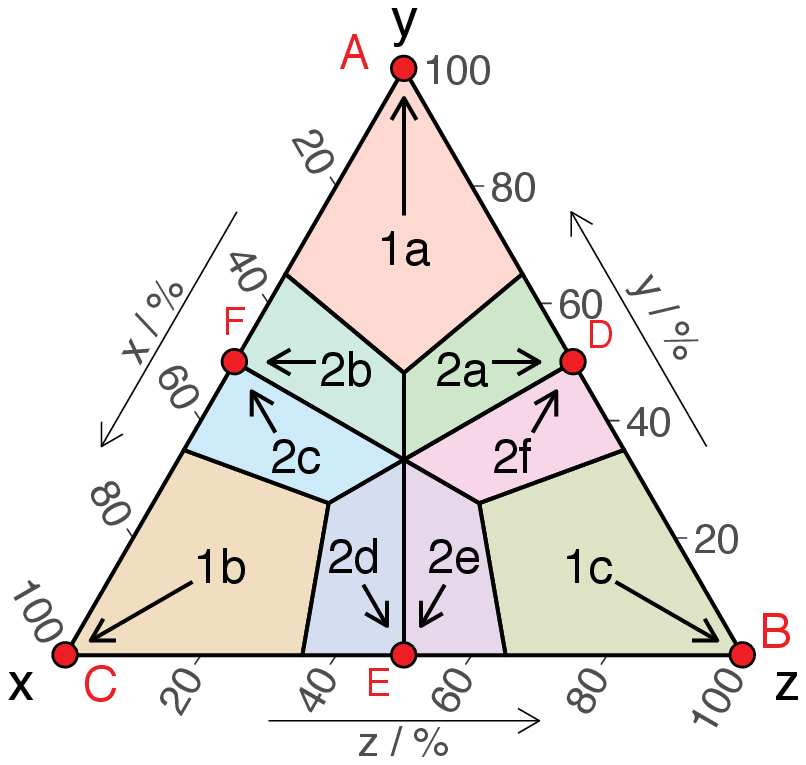}
	}
	\subfloat[$p=0.65$\label{fig:trans_prob_regions_p_0.65}]{%
	\includegraphics[trim=0.35in 0.41in 0.43in 0.3in,clip,width=0.5\columnwidth]{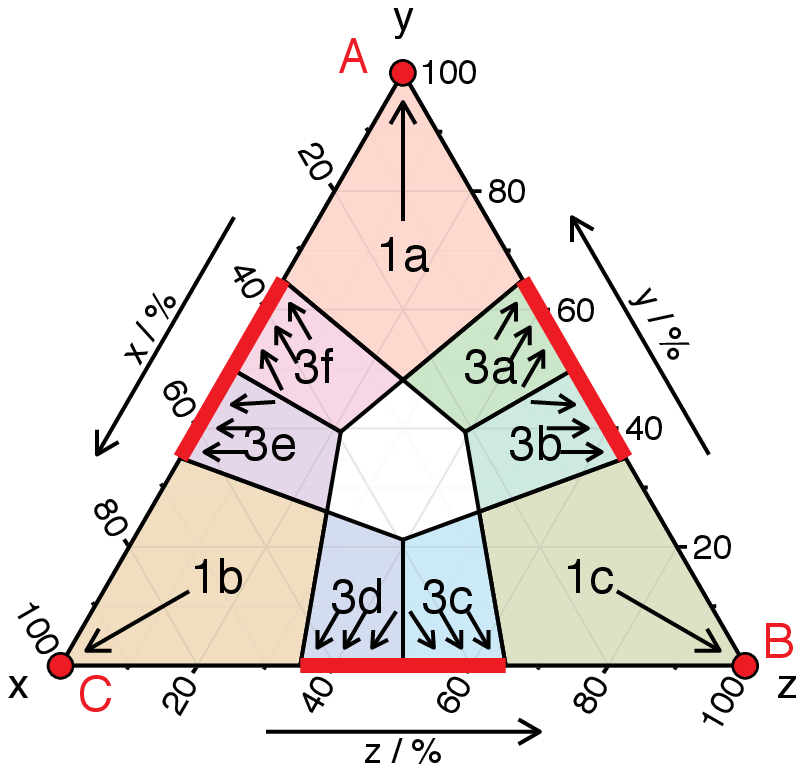}	
	}
	
\protect\caption{\label{fig:trans_prob_regions}
Attractors and their corresponding basins of attraction (by the direction of the arrows in the region) of Eq.~\ref{eq:master_eq_short} in the configuration space when \protect\subref{fig:trans_prob_regions_p_0.35} $p=0.35$ and when \protect\subref{fig:trans_prob_regions_p_0.65} $p=0.65$, in the thermodynamic limit.
Point A is the attractor of region 1a and point F is the attractor of both region 2b and region 2c in \protect\subref{fig:trans_prob_regions_p_0.35}.
In \protect\subref{fig:trans_prob_regions_p_0.65}, the attractors for regions 3a to 3f are the red thick intervals marked on the edges of the configuration space.
}
\end{figure}

First we consider systems of very large size, i.e., in the thermodynamic limit. 
Here we use $x=N_1/N,y=N_2/N,z=N_3/N$ instead of $N_1,N_2,N_3$. 
Please refer to Fig.~\ref{fig:trans_prob_regions} for attractors and their corresponding basins of attraction of Eq.~\ref{eq:master_eq_short} in the configuration space.
The derivations of the boundaries of each basins of attraction can be found in the appendix~A-E.
Here we emphasize that the boundaries depends only on $p$ in the thermodynamics limit.
Any interior point in the configuration space represents a fragmented state.
The time evolution of the system can be represented as a path in the configuration space.
Once the point enters the edges of the configuration space, implying that the population is in polarized state, it cannot go back into the interior of the configuration space.
According to \cite{benczik_opinion_2009}, when $p<0.5$, the relaxation time scale from meta-stable states, D, E, F, increases exponentially with the population size $N$.
The three vertices of the configuration space, i.e., points A, B and C, imply that the population reach the consensus state.
Therefore, points A, B and C are the absorbing states of the system.
The union of regions 1a, 1b and 1c, which brings the system to the consensus state A, B, or C, is denoted by $\mathbb{B}$.

Now suppose the population density $P(x,y)$ is a two-dimensional delta function centered on $(x_0,y_0)$.
Denote by $(x(t),y(t))$ the trajectory the population will follow \emph{spontaneously} according to Eq.~\ref{eq:master_eq_short}.
Refer to Fig.~\ref{fig:trans_prob_regions_p_0.35} again ($p=0.35$).
In region 1a, $x(t)=x_0 e^t$ and $y(t)=(y_0-1) e^{-t}+1$.
Similarly, in regions 1b and 1c, the probability mass will approach the attractors in $e^{-t}$ fashion.
In region 2a, $x(t)=x_0 e^{-t}$ and $y(t)=1/2-(1/2-y_0) e^{-2t}$.
Therefore the probability mass will approach the attractor also in $e^{-t}$ fashion.
In region 2f, $x(t)=x_0 e^{-t}$ and $z(t)=1/2-(1/2-z_0)e^{-2t}$.
For this reason, though regions 2a and 2f share the same attractor, we still mark them as different basins of attraction.
In summary, in the thermodynamics limit, when $p<0.5$, the fragmented state cannot sustain itself, any populations that are within regions 2a to 2f will be stuck in the polarized state and any populations that are inside regions 1a to 1c will go directly from fragmented state to consensus state.
In two-opinion system, according to \cite{benczik_opinion_2009}, populations with initial condition $pN<N_1<(1-p)N$ (if $p<0.5$) will be attracted to the \textit{totally polarized state}, i.e., $N_1=N_2=N/2$, which is a meta-stable state.
If $p>0.5$, populations with initial condition $(1-p)N<N_1<pN$ will be frozen.
Therefore, in order for the population to ever reach the consensus state in thermodynamic limit, one has to deliberately convert some of the opinion holders to change the distribution of different opinions, which requires significant effort.
Given an initial opinion distribution $\bold{X}_0=(x_0,y_0,z_0)$ and the desired opinion distribution $\bold{X}_1=(x_1,y_1,z_1)$, one can describe the conversion plan by six non-negative numbers: $\Delta_{1\rightarrow2}$, $\Delta_{1\rightarrow3}$, $\Delta_{2\rightarrow1}$, $\Delta_{2\rightarrow3}$, $\Delta_{3\rightarrow1}$ and $\Delta_{3\rightarrow2}$, where $\Delta_{i\rightarrow j}$ means the fraction of opinion $i$ holders needed to be converted to opinion $j$ holders.
It is reasonable to measure the effort in terms of
\begin{equation}
\begin{aligned}
	\Delta_{1\rightarrow2}+\Delta_{1\rightarrow3}+\Delta_{2\rightarrow1}+\Delta_{2\rightarrow3}+\Delta_{3\rightarrow1}+\Delta_{3\rightarrow2},\\
	\text{s.t. }\Delta_{2\rightarrow1}+\Delta_{3\rightarrow1}-\Delta_{1\rightarrow2}-\Delta_{1\rightarrow3}=x_1-x_0,\\
	\Delta_{1\rightarrow2}+\Delta_{3\rightarrow2}-\Delta_{2\rightarrow1}-\Delta_{2\rightarrow3}=y_1-y_0,\\
	\Delta_{1\rightarrow3}+\Delta_{2\rightarrow3}-\Delta_{3\rightarrow1}-\Delta_{3\rightarrow2}=z_1-z_0.
\label{eq:intervention_cost}
\end{aligned}
\end{equation}
One can show that (refer to appendix F) the least fraction of population affected, given initial opinion distribution $\bold{X}_0$ and the desired opinion distribution $\bold{X}_1$, is
\begin{equation}\label{eq:min_cost}
	\delta n_3(\bold{X}_0,\bold{X}_1)\equiv\underset{\{\Delta_{i\rightarrow j}\}}{\text{min}}\sum_{i\neq j,i,j=1}^{3}\Delta_{i\rightarrow j}=\frac{1}{2}\|\bold{X}_1-\bold{X}_0\|_1,
\end{equation}
which we will call the \textit{minimal intervention cost} and $\|\cdot\|_1$ is the $L1$-norm.
This measure can be directly generalized to all numbers of available opinions, i.e., $\delta n_c$ where $c=2,3,4,\cdots$.
For the derivation of the measure, please refer to appendix F.

\begin{figure}
	
	\subfloat[$p<0.5$\label{fig:hitchhike_p_0.2}]{%
	\includegraphics[trim=0.35in 0.4in 0.43in 0.3in,clip,width=0.7\columnwidth]{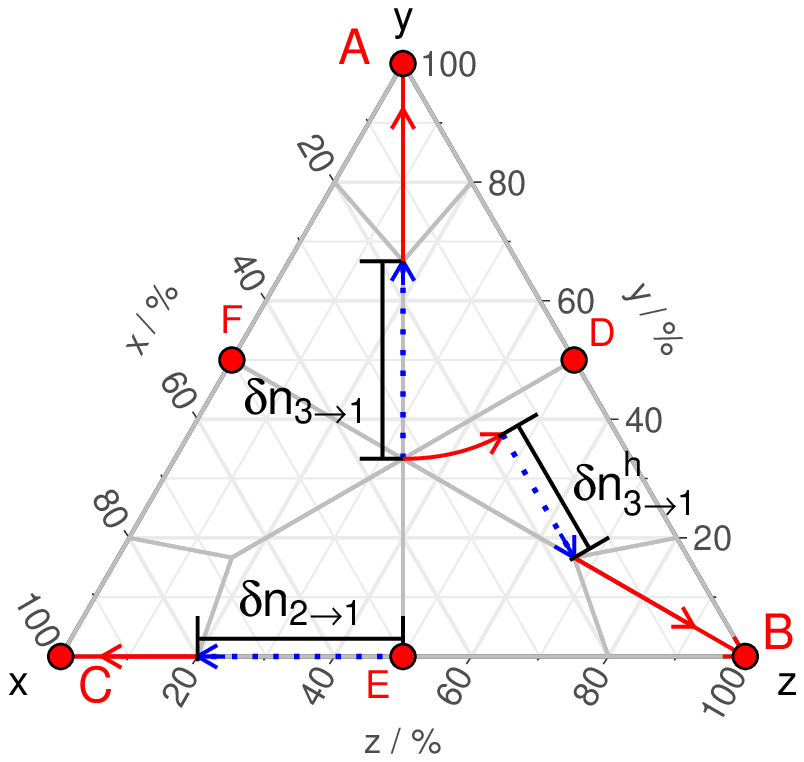}
	}\\
	\subfloat[$p>0.5$\label{fig:hitchhike_p_ge_0.5}]{%
	\includegraphics[trim=0.35in 0.4in 0.43in 0.3in,clip,width=0.7\columnwidth]{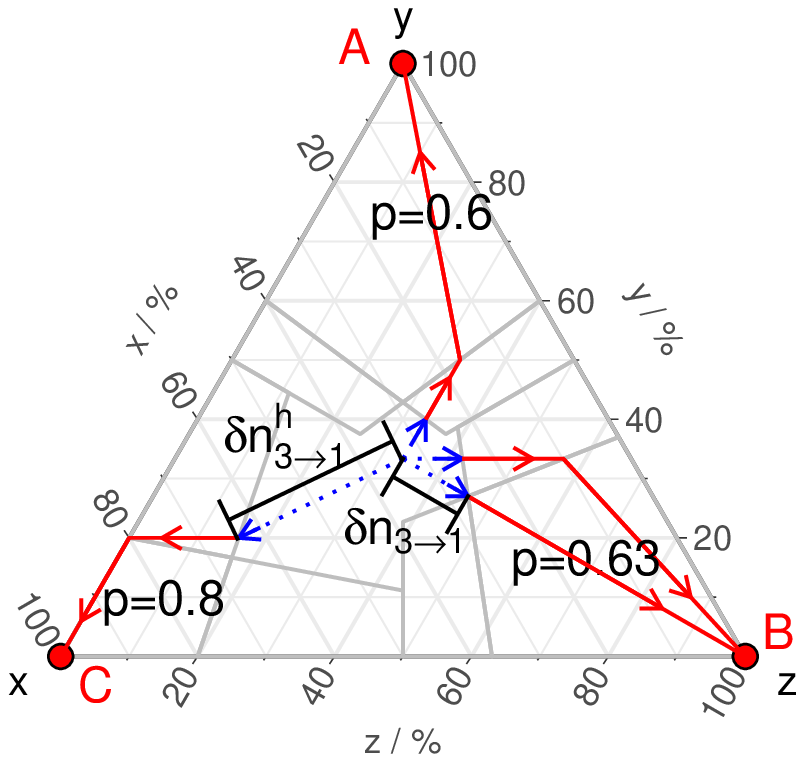}
	}
	
\protect\caption{\label{fig:hitchhike}
Various paths of opinion conversions and corresponding minimal intervention cost shown in the configuration space.
The blue dotted lines with arrow heads represent the deliberate conversion, with half its $L1$-norm being the minimal intervention cost (as defined in Eq.~\ref{eq:min_cost}).
The red solid lines represent spontaneous dynamics.
(a) When $p<0.5$, the path of \emph{naive opinion conversion} (defined in Eq.~\ref{eq:naive}) is shown from the center, $\mathbb{U}_0$, to attractor A, and its minimal intervention cost is measured by $\delta n_{3\rightarrow1}$.
However, as shown by the path from the center to attractor B, we can hitchhike on the spontaneous dynamics by first allowing it to evolve spontaneously, following the path of $\bold{X}_{2a}(t)$ in Eq.~\ref{eq:hitchhike}. 
Then we make a deliberate conversion, illustrated by the blue dotted segment, its minimal conversion cost measured by $\delta n_{3\rightarrow1}^h$.
According to Eq.~\ref{eq:min_cost}, we end up spending less effort (or shorter time) in pushing the population to the consensus state, or attractor B. 
Note that this intervention cost, $\delta n_{3\rightarrow1}^h$ is even smaller than that of a two-opinion system to attractor C, measured by $\delta n_{2\rightarrow1}$, its conversion path from $(0.5,0,0.5)$ to attractor C.
(b) When $p>0.5$, the boundaries of various attractors under different $p$ are shown in the same figure. 
Naive conversions, such as the the path from the center directly to attractor B, cost more effort (measured by $\delta n_{3\rightarrow1}$) to reach the consensus state.
The nearby path to the same attractor B takes less intervention effort by entering region 3b (shaded region), and the spontaneous dynamics will send the system to attractor B.
The path from $\mathbb{U}_0$ to attractor C shows that when $p>0.8$, the minimal intervention cost by hitchhiking, measured by $\delta n_{3\rightarrow1}^h$ is almost as large as that by naive deliberate conversion, measured by $\delta n_{3\rightarrow1}$.
The path from $\mathbb{U}_0$ to attractor A is shown here to emphasize the effect of $p$ on the paths with the minimal conversion cost.
}
\end{figure}

In particular, we are interested in the minimal intervention cost necessary to persuade the population with three initial opinions at the totally fragmented/polarized state to reach the consensus state.
One does not need to persuade the population all the way from fragmented state to the consensus state.
Instead, it is sufficient to push the population into the edge of $\mathbb{B}$ region.
The minimal intervention cost in this case, which will be called the \emph{naive conversion}, is
\begin{equation}\label{eq:naive}
	\delta n_{3\rightarrow1}=\min_{\bold{X}_a\in\mathbb{B}}\delta n_3(\bold{U}_0,\bold{X}_a),
\end{equation}
where $\mathbb{B}$ is the union of the basins of attraction of attractors A, B and C in Fig.~\ref{fig:trans_prob_regions} and $\bold{U}_0=(\frac{1}{3},\frac{1}{3},\frac{1}{3})$ is the totally fragmented state.
For two-state system, $\delta n_{2\rightarrow1}=\frac{1}{2}-p$ for $p<0.5$ (indicated in Fig.~\ref{fig:hitchhike_p_0.2}) and $\delta n_{2\rightarrow1}=p-\frac{1}{2}$ for $p>0.5$.
For three-state system, $\delta n_{3\rightarrow1}=\frac{1 - p}{p + 1} - \frac{1}{3}$ for $p<0.5$ and $\delta n_{3\rightarrow1}=\frac{p}{2 - p} - \frac{1}{3}$ for $p>0.5$ (indicated in Fig.~\ref{fig:hitchhike}).

A unique feature of three-state system is that one can exploit (or ``hitchhike'' on) the \emph{spontaneous dynamics} in regions 2a-2f and 3a-3f in Fig.~\ref{fig:trans_prob_regions} and the geometry of neighboring region to minimize the effort spent in deliberately changing the opinion distribution.
To capture the reduction in effort made possible by hitchhiking, we denote the minimal intervention cost by hitchhiking by $\delta n_{3\rightarrow1}^h$.
When $p<0.5$, we have
\begin{equation}\label{eq:hitchhike}
	\delta n_{3\rightarrow1}^h=\min_{t\in(0,\infty),\bold{X}_a\in\mathbb{B}}\delta n_3(\bold{X}_{2a}(t),\bold{X}_a),
\end{equation}
where $\bold{X}_{2a}(t)=(\frac{1}{3}e^{-t},\frac{1}{2}-\frac{1}{6} e^{-2t},\frac{1}{2}-\frac{1}{3}e^{−t}+\frac{1}{6}e^{−2t}$), which is just the path a Dirac-delta probability mass will follow according to Eq.~\ref{eq:master_eq_short}.
Fig.~\ref{fig:hitchhike_p_0.2} shows how one can ``hitchhike'' on the dynamics to spend less effort in pushing the population to the consensus state.

When $p>0.5$, hitchhiking on the spontaneous dynamics yields
\begin{equation}
\begin{aligned}
	&\delta n_{3\rightarrow1}^h=\min_{\bold{X}_a\in\mathbb{C}}\delta n_3(\bold{U}_0,\bold{X}_a)\\
	&\text{s.t. }\lim_{t\rightarrow\infty}\bold{X}(t|\bold{X}_a)=(1,0,0)\text{ or }(0,1,0)\text{ or }(0,0,1),
\end{aligned}
\end{equation}
where $\mathbb{C}$ is the union of all colored regions in Fig.~\ref{fig:trans_prob_regions_p_0.65}.
Fig.~\ref{fig:hitchhike_p_ge_0.5} shows how hitchhiking is accomplished in the configuration space, and $\bold{X}(t|\bold{X}_a)$ is the trajectory of Dirac-delta probability mass once it enters $\mathbb{C}$.

\begin{center}
\begin{figure}
\begin{centering}
\includegraphics[trim=0.2in 0in 0.2in 0in,clip,width=0.9\columnwidth]{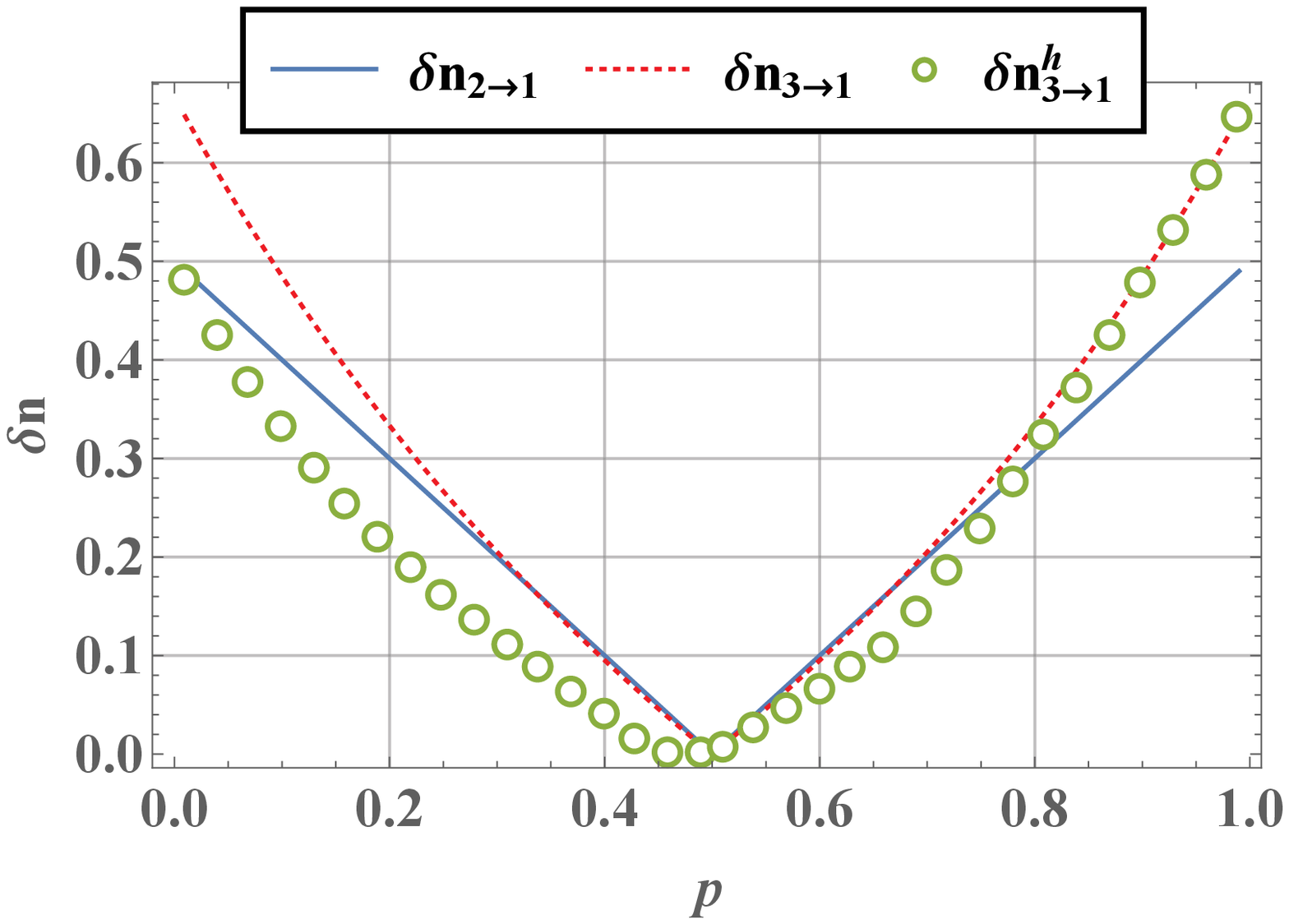}
\par\end{centering}

\protect\caption{\label{fig:minimal_fraction_converted}
Minimal intervention cost to reach the consensus state as a function of $p$.
Hitchhiking on the spontaneous dynamics in three-state system (the effort is measured by $\delta n_{3\rightarrow1}^h$, the green line in the figure) in order to reach the consensus state requires less effort compared to naively pushing the population to the nearby consensus state attractor (points A, B, C) (the effort is measured by $\delta n_{3\rightarrow1}$) for different values of $p$.
Also $\delta n^h_{3\rightarrow1}<\delta n_{2\rightarrow1}$ for $0<p<0.8$.
}

\end{figure}

\par\end{center}

The calculation of minimal intervention cost can be found in appendix G.
In Fig.~\ref{fig:minimal_fraction_converted} we show the minimal intervention cost required to lift a population out of totally fragmented/polarized state.
The smaller the values, the easier for the population to reach the consensus state.
While naively pushing the population to the nearest consensus state attractors in three-state systems always requires larger effort than in two-state system, hitchhiking on the spontaneous dynamics in three-state system to reach the consensus state actually lowers the minimal effort compared to that in two-state systems, when $p<0.8$.
This is partly due to the fact that there is more room to intervene in a two-dimensional configuration space than in a one-dimensional space.
Therefore, we conclude that it is easier, or requires less intervention, to reach the consensus state in three-state system than in two-state system for a broad range of $p$.

\section{Finite size effect}

In finite systems, a polarized/fragmented population may still spontaneously reach the consensus state due to stochastic drift.
We conduct Monte Carlo simulation of a finite system with totally polarized/fragmented initial condition to investigate the finite size effect.
We also perform numerical integration of Eq.~\ref{eq:master_eq_short} to investigate the time evolution of $P(x,y,t)$.

\begin{figure}
	\subfloat[$p=0.29$.]
	{\includegraphics[trim=1in 1in 1.9in 1in,clip,width=0.3\columnwidth]{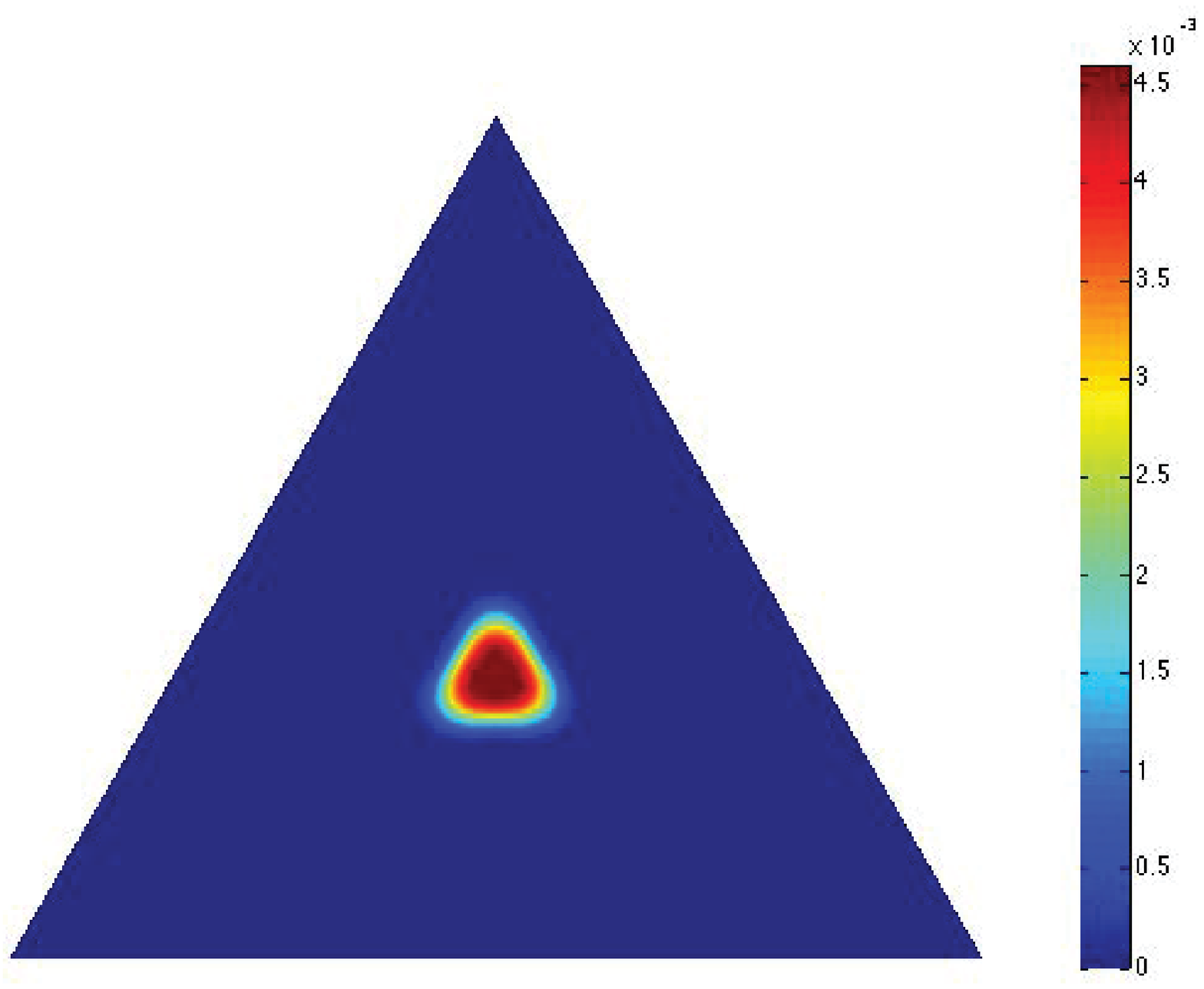}	
	\includegraphics[trim=1in 1in 1.9in 1in,clip,width=0.3\columnwidth]{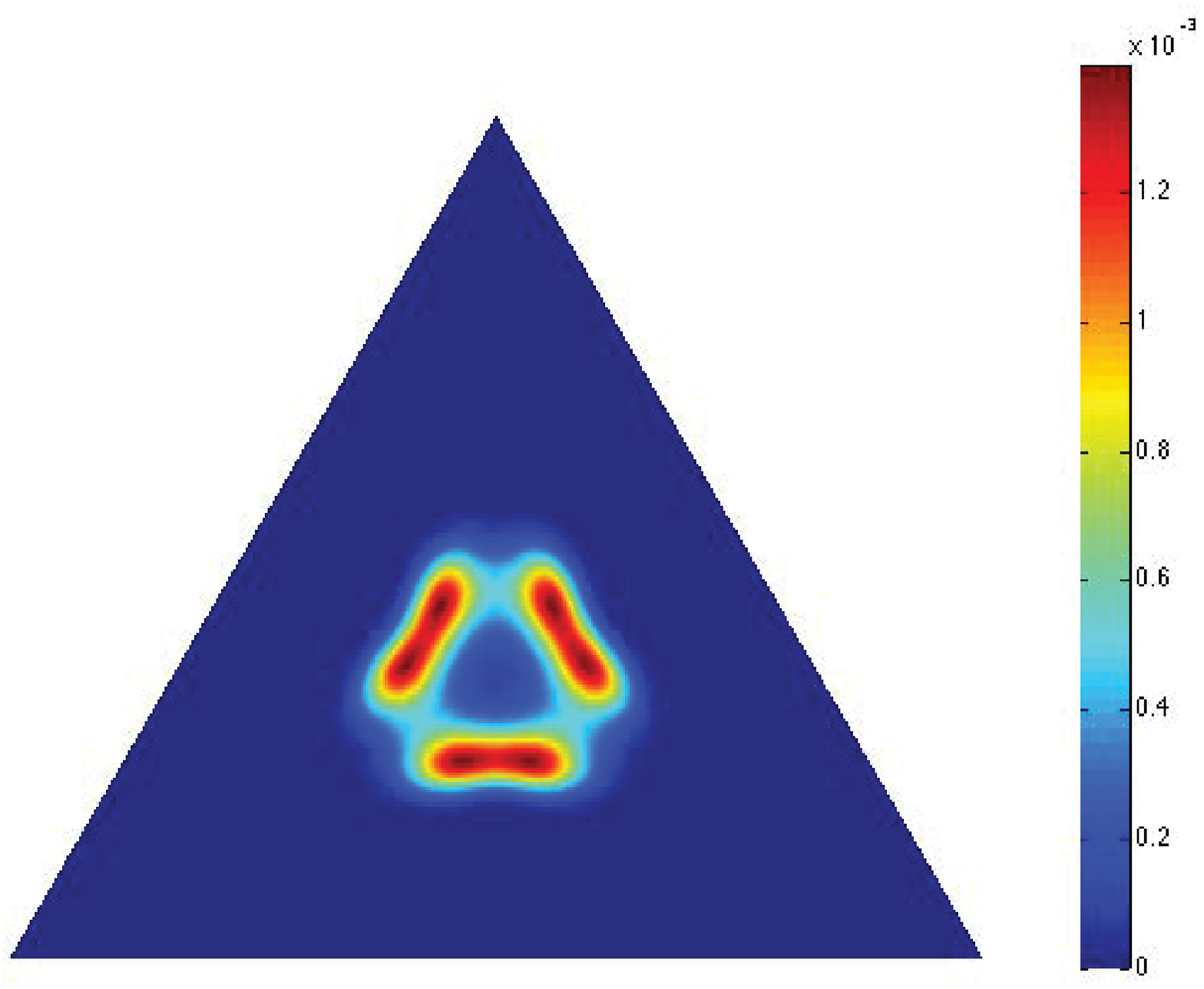}
	\includegraphics[trim=1in 1in 1in 1in,clip,width=0.36\columnwidth]{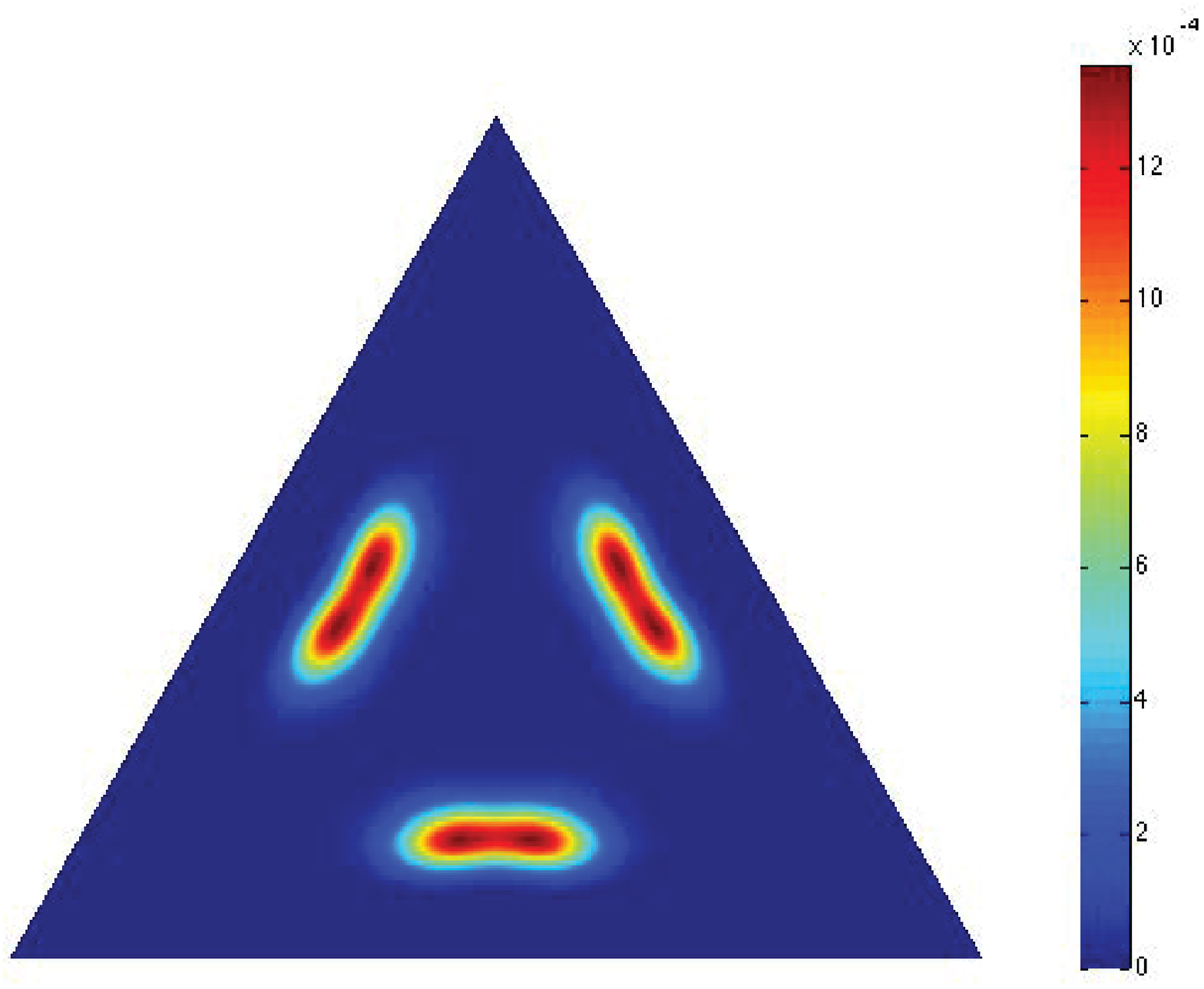}}\\
	\subfloat[$p=0.4$.]
	{\includegraphics[trim=1in 1in 1.9in 1in,clip,width=0.3\linewidth]{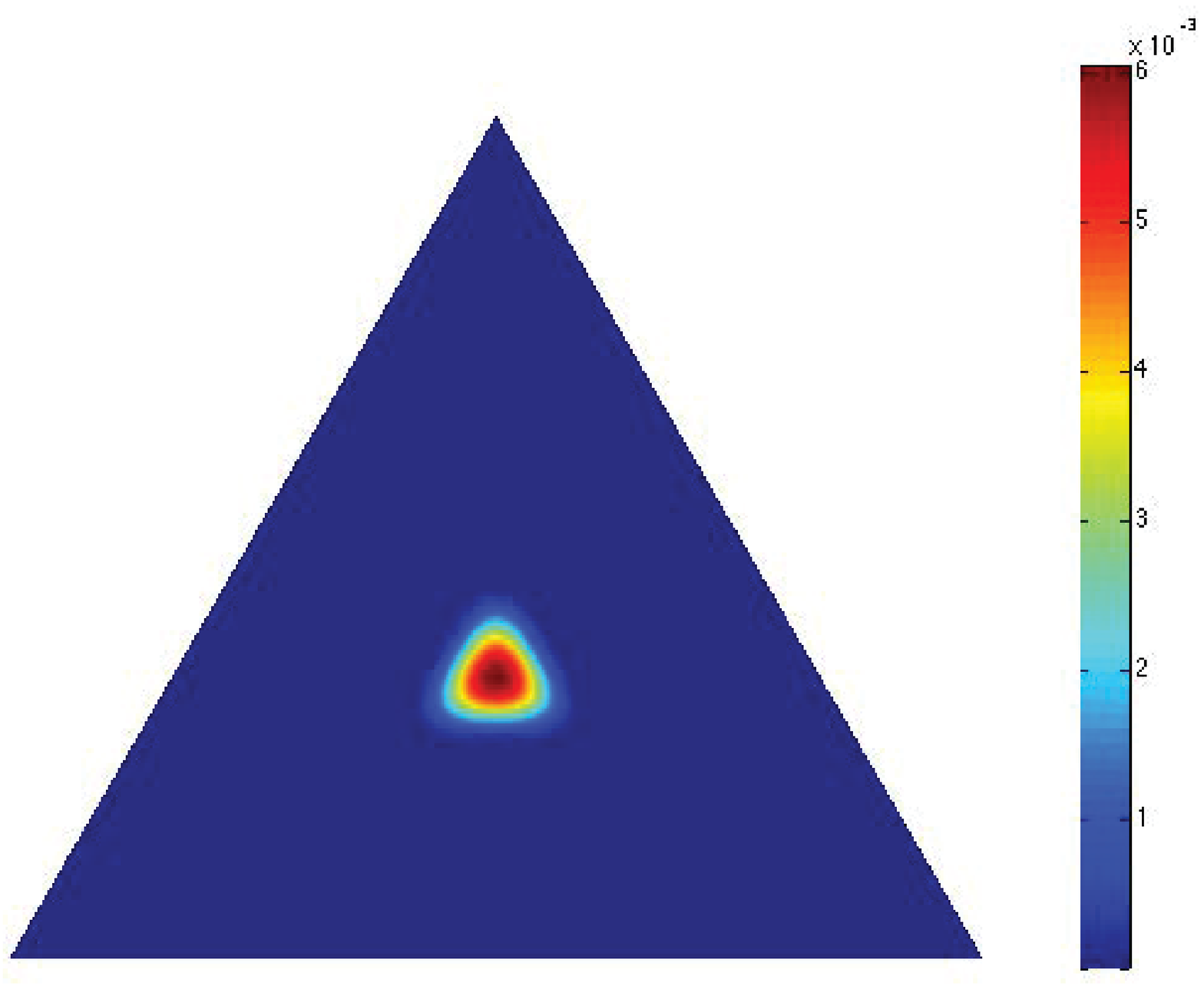}	
	\includegraphics[trim=1in 1in 1.9in 1in,clip,width=0.3\linewidth]{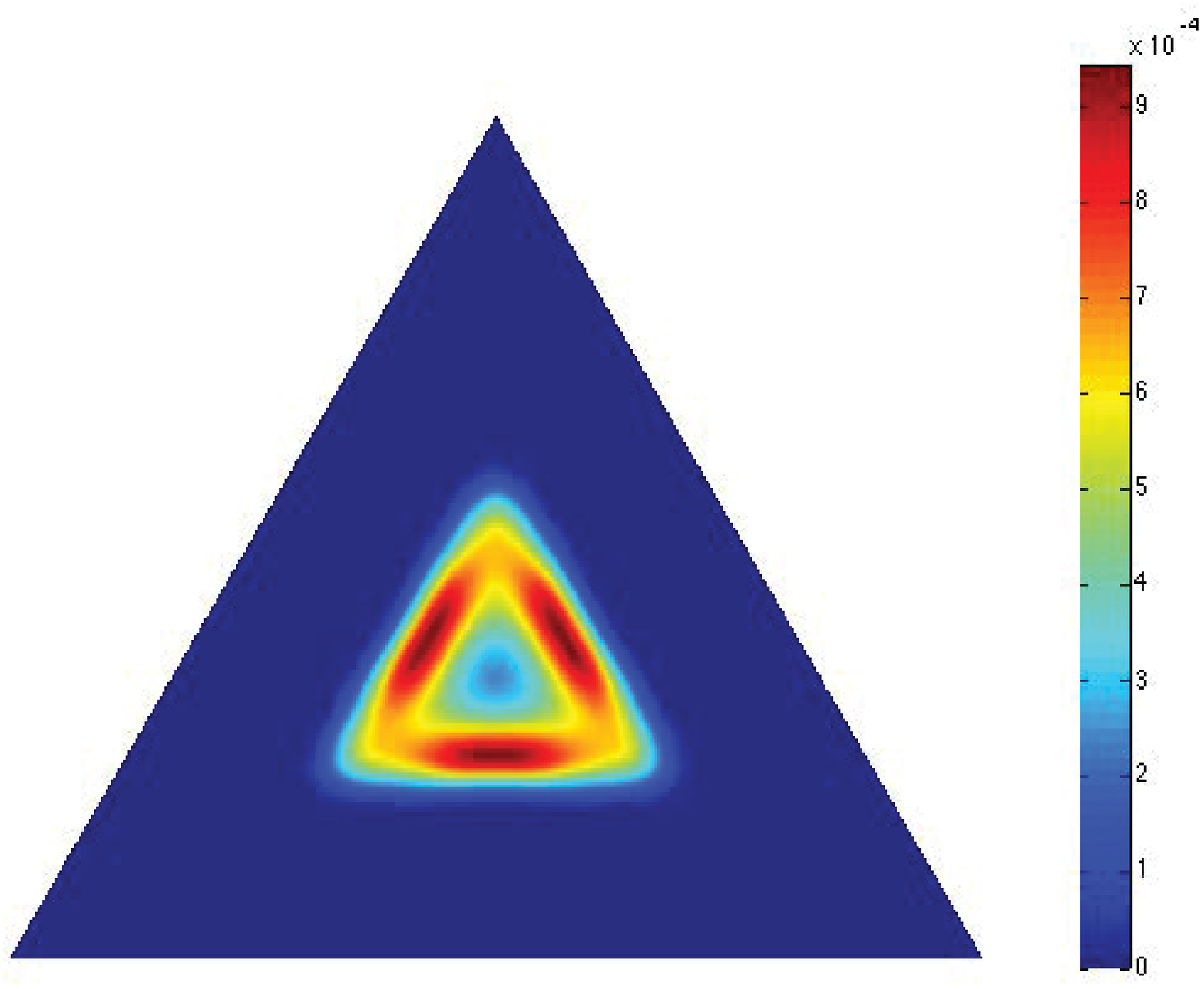}
	\includegraphics[trim=1in 1in 1in 1in,clip,width=0.36\linewidth]{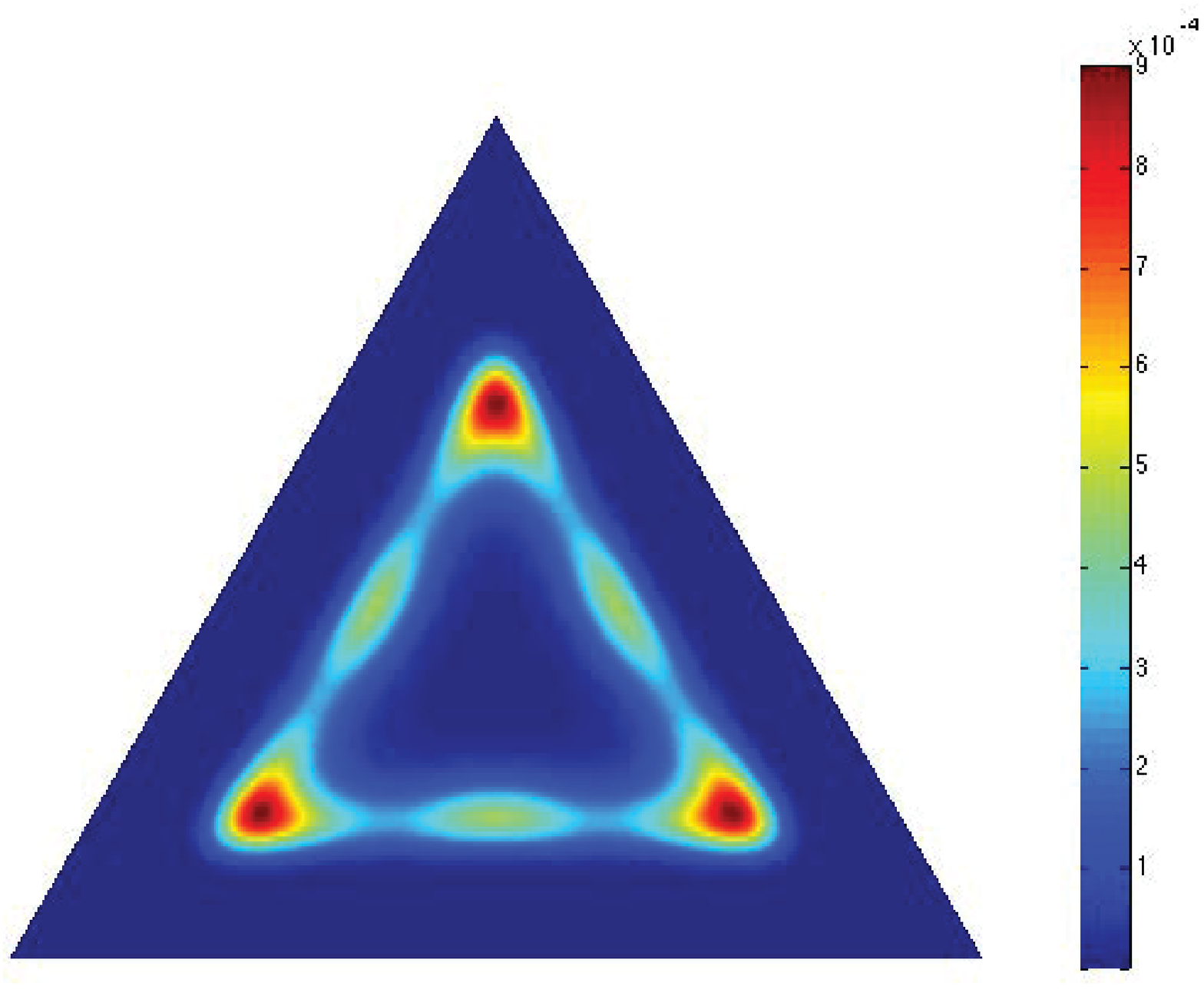}}	\\
	\subfloat[$p=0.5$.]
	{\includegraphics[trim=1in 1in 1.9in 1in,clip,width=.3\linewidth]{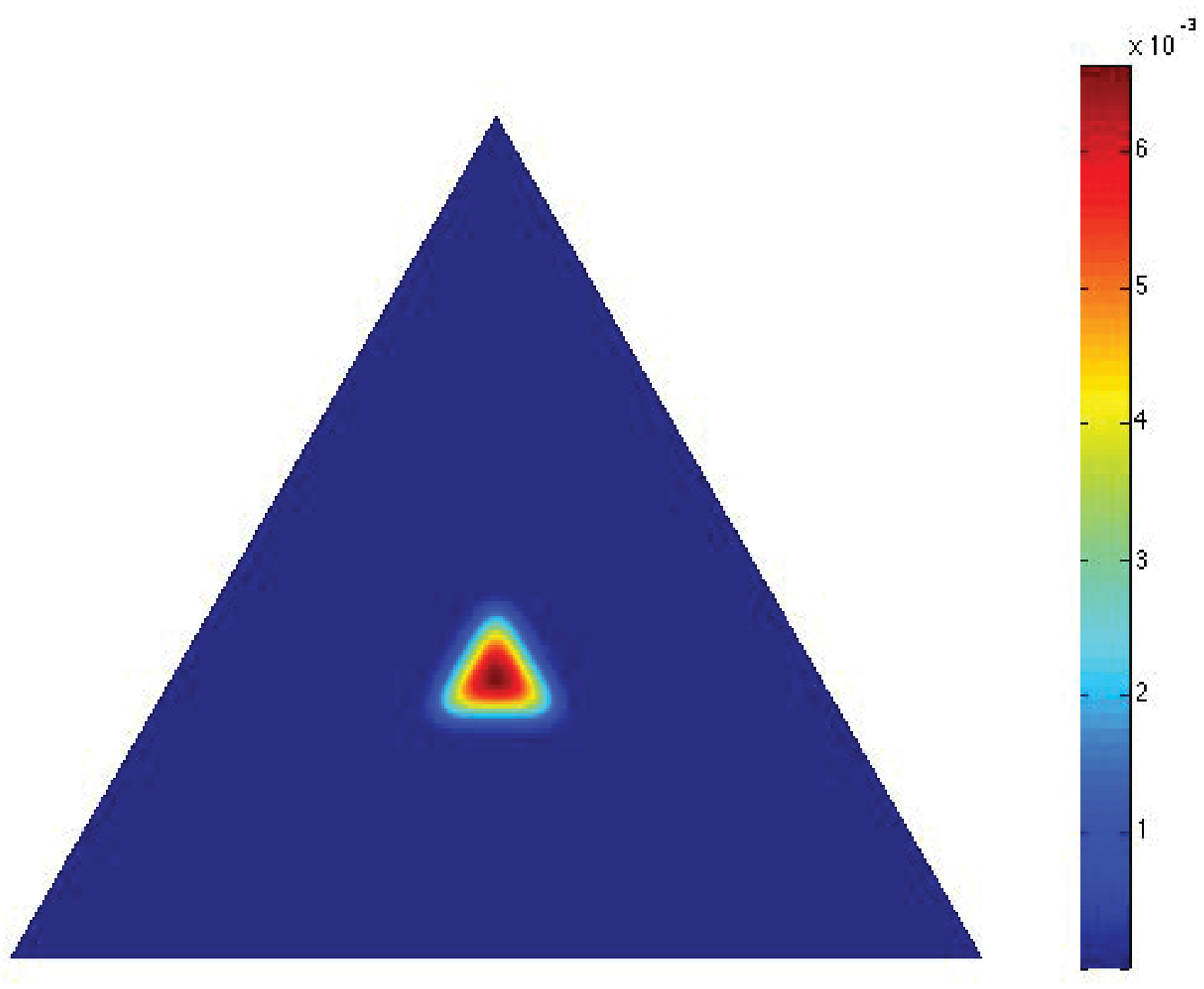}	
	\includegraphics[trim=1in 1in 1.9in 1in,clip,width=.3\linewidth]{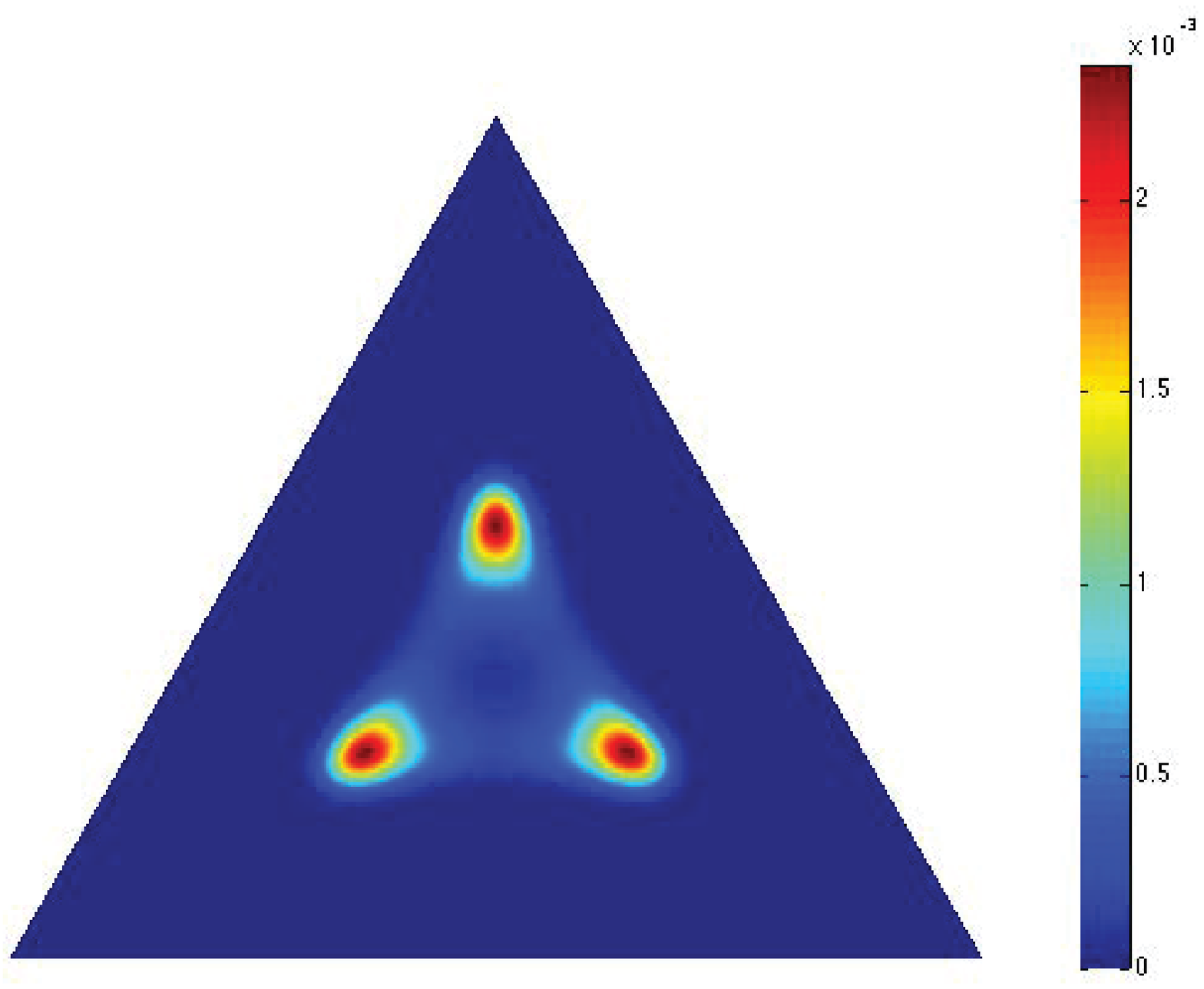}	
	\includegraphics[trim=1in 1in 1in 1in,clip,width=.36\linewidth]{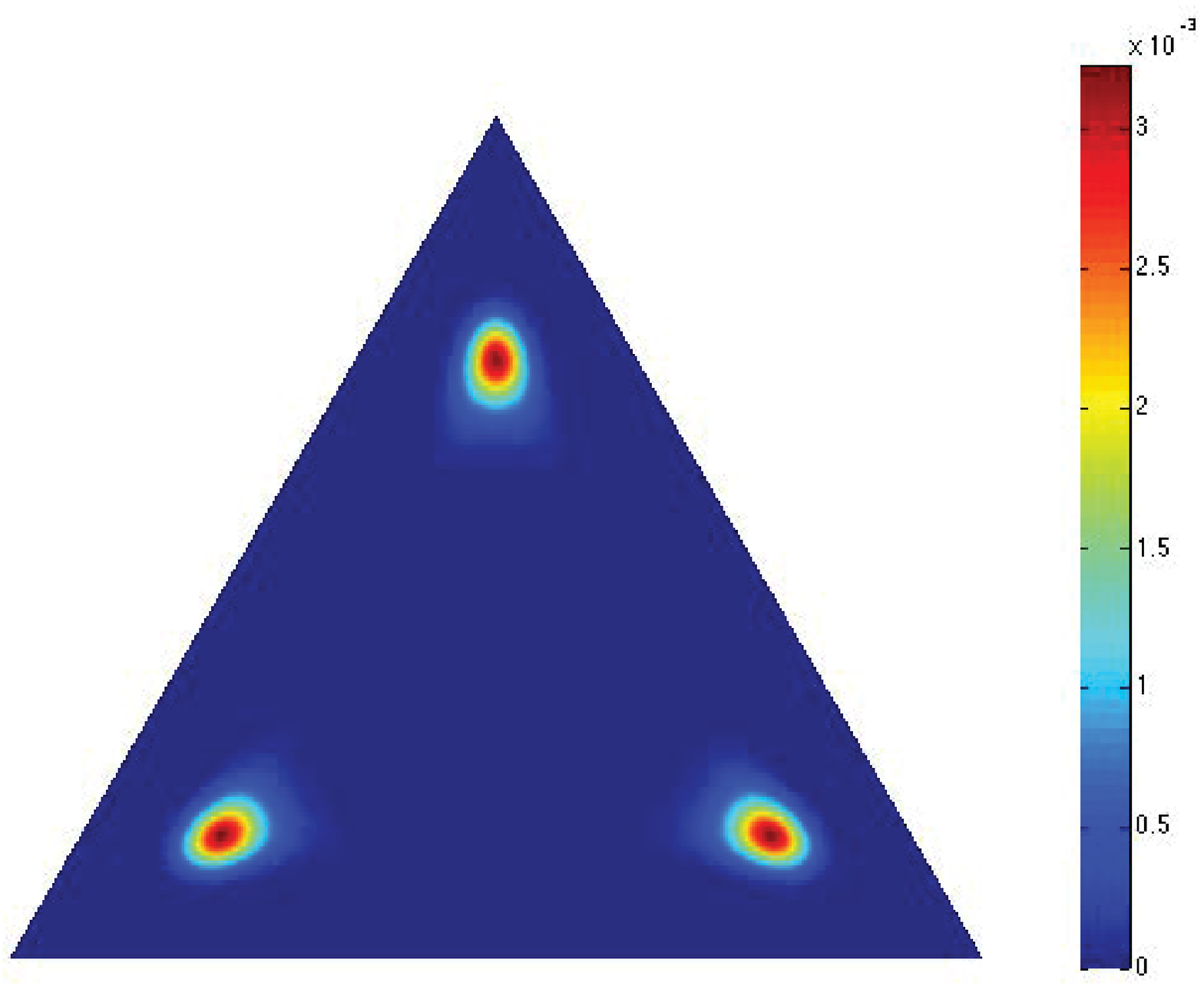}}

\protect\caption{\label{time_evo_N_150}Time evolution of $P(x,y,t)$ obtained by numerically integrating Eq.~\ref{eq:master_eq_short}. 
The snapshots are taken (from left to right) at time $t=0.2$, $t=0.4$ and $t=1$.
$N=150$.
See \cite{OF3_supp} for animations depicting the evolutions of $P(x,y,t)$ when $p=0.29$, $0.4$ or $0.5$.
}

\end{figure}

In Fig.~\ref{time_evo_N_150} we show several snapshots of  $P(x,y,t)$ with $p\le0.5$ and $N=150$.
The snapshots clearly reflect the existence of the six attractors and their basins of attraction depicted in Fig.~\ref{fig:trans_prob_regions_p_0.35}.
In the beginning, the Kronecker Delta probability mass $P(x,y,t=0)=\delta_{x,1/3}\delta_{y,1/3}$ diffuses into a distribution with finite width, due to finite size effect.
When $p=0.29$, the finite size effect forces all of the probability mass into regions~2a to 2f.
When $p=0.35$, considerable amount of probability mass enters regions~1a to 1c.
When $p=0.5$, regions~2a to 2f no longer play a role in the time evolution.

\begin{center}
\begin{figure}
\begin{centering}
\includegraphics[width=0.8\columnwidth]{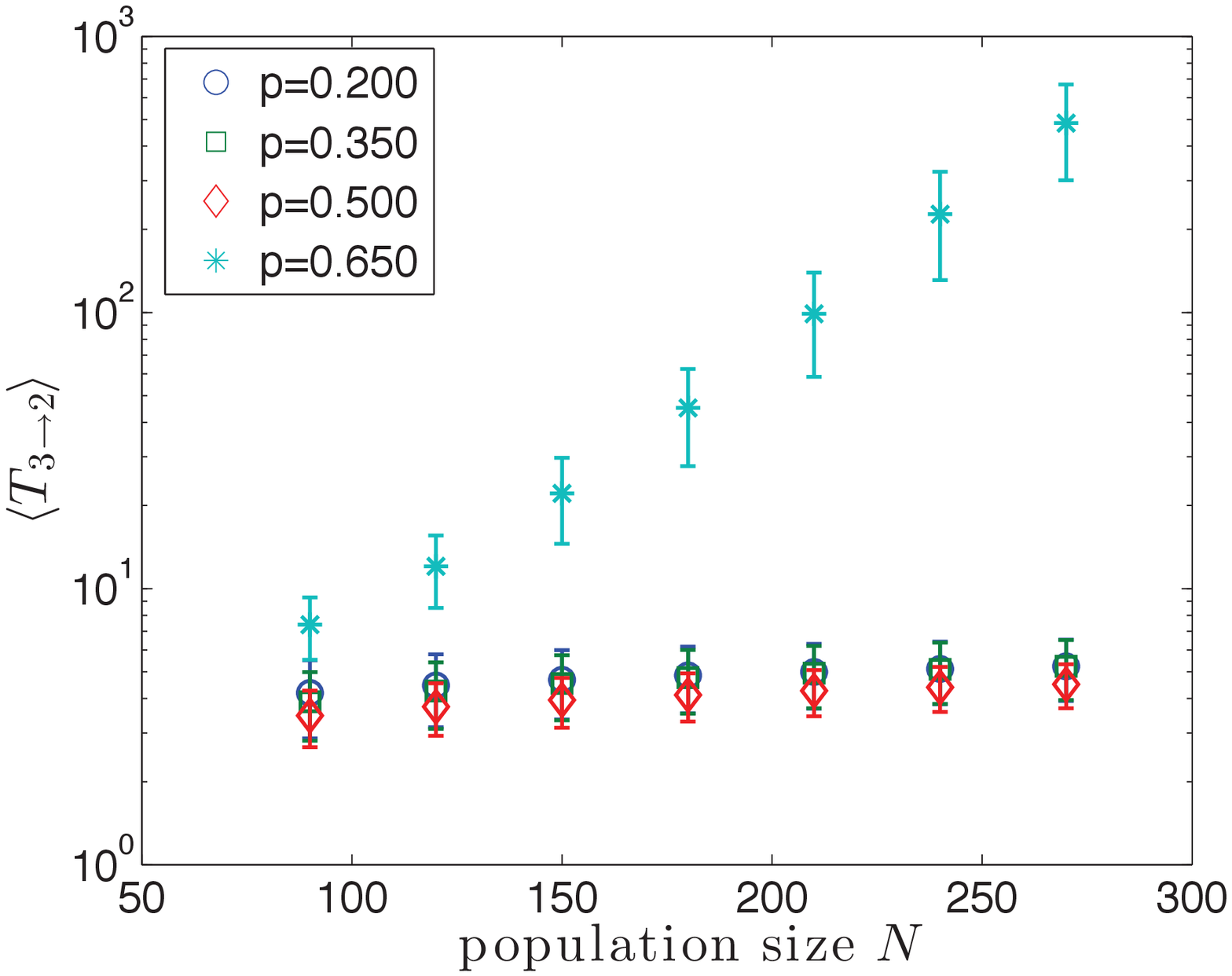}
\par\end{centering}

\protect\caption{\label{fig:avg_T_3to2}$\left<T_{3\rightarrow2}\right>$ as a function
of $N$ for various values of $p$. The error bars are the standard
deviations of $\left<T_{3\rightarrow2}\right>$. $\left<T_{3\rightarrow2}(p=0.65)\right>\sim\exp(0.026N)$.}

\end{figure}

\par\end{center}

The finite size effect can be investigated by Monte Carlo simulation with a random sequential updating scheme.
In each Monte Carlo step, an agent $i$ is randomly selected. 
We consider all pairs of agent, ($i,j$), independently. 
If $\sigma_{i}=\sigma_{j}$, the pair links with probability $p$; otherwise they link with probability $q$. 
Once all choices are made, $\sigma_{i}$ is updated following a plurality rule: if there exists a plurality opinion $\alpha^{*}$ such that
\begin{equation}
N_{\alpha^{*}}>N_{\beta}\quad\forall\beta\neq\alpha^{*},
\end{equation}
then we assign $\sigma_{i}=\alpha^{*}$. 
Here $N_{\alpha}$ is the number of opinion $\alpha$ in the neighborhood. 
Otherwise $\sigma_i$ is unchanged. 
The temporary linking information will be discarded after their updating procedure before we proceed to the next Monte Carlo step. 
In our analysis, one unit of time corresponds to $N$ Monte Carlo steps. 

To analyze the mechanism of the finite size effect, let us break down the consensus reaching process of finite system into two subprocesses: 
\begin{itemize}
\item Process \rom{1}: one of the three opinions goes extinct. 
\item Process \rom{2}: the population with the two remaining opinions finally reaches the consensus state. 
\end{itemize}
The final state of process \rom{1} is a distribution of the three opinions $(x,y,z)$, which is also the initial state of process \rom{2}.
By virtue of symmetry and without loss of generality, we can lump several final states into one state, so as to simplify the notations.
We define the lumped final state $m$ as a set of configurations and the shorthand $m=a$ means
\begin{equation}
\begin{aligned}
	&m= \\
	&\left\{ (0,a,N-a),(0,N-a,a),(a,0,N-a),\right.\\
	&\left.(N-a,0,a),(a,N-a,0),(N-a,a,0) \right\}.
\end{aligned}
\end{equation}
Therefore, we denote the duration of process \rom{1} as $T_{3\rightarrow2,m}$, where $m$ is the final state of process \rom{1} defined above.
Similarly, the duration of process \rom{2} is denoted by $T_{2\rightarrow1|m}$, where $m$ indicates the initial state of process \rom{2}. 

Now, we introduce some more notations:
$P_{2\rightarrow1|m}(T)$ is the distribution of $T_{2\rightarrow1|m}$ and $P_{3\rightarrow2,m}(T)$ is the distribution of $T_{3\rightarrow2,m}$.
Therefore, the distribution of the duration of the whole process, $T_{3\rightarrow1}\equiv T_{3\rightarrow2,m}+T_{2\rightarrow1|m}$, is
\begin{equation}
\label{eq:T3_1_dist}
P_{3\rightarrow1}(T)=\sum_{a=0}^{N/2} \sum_{t=0}^T P_{3\rightarrow2,a}(T-t)P_{2\rightarrow1|a}(t)P_m(a),
\end{equation}
where $P_m(a)$ is the probability that $m=a$ at the end of process \rom{1}.
We define $\left<T_{3\rightarrow2}\right>=\sum_{a=0}^{N/2} \sum_{T=0}^\infty P_{3\rightarrow2,a}(T)TP_m(a)$. 
Fig.~\ref{fig:avg_T_3to2} shows that, if $p<0.5$, $\left<T_{3\rightarrow2}\right>$ is insensitive to $N$ and $p$, which is consistent with our analysis of large systems above.
Therefore, we will routinely make the approximation that $\left<T_{3\rightarrow2,m}\right>\approx\left<T_{3\rightarrow2}\right>$ when $p<0.5$.
If $p>0.5$, $\left<T_{3\rightarrow2}\right>$ scales as $\exp(N)$. 

\begin{center}
\begin{figure}[h]
\begin{centering}
\includegraphics[trim=0in 0in 0in 0in,clip,width=\columnwidth]{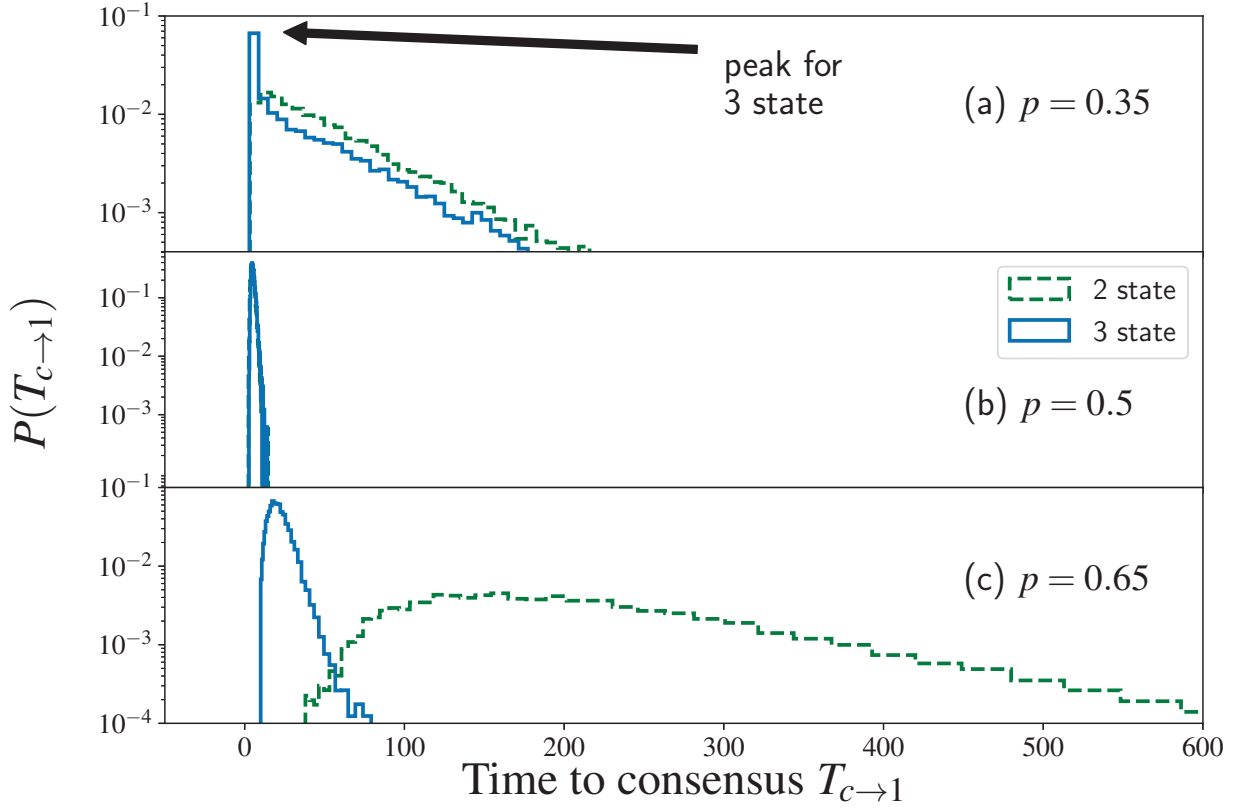}
\par\end{centering}

\protect\caption{\label{fig:Tc1dist_c_2_3}
Distribution of time to consensus for two-state and three-state system obtained from Monte Carlo simulation.
The green line is the two state data, $T_{2\rightarrow1}$, while the blue dotted line is for the three state model, $T_{3\rightarrow1}$.
$N=150$.
Note that when $p=0.35$, there is a sharp peak in the distribution of $T_{3\rightarrow1}$.
}

\end{figure}
\end{center}

In Fig.~\ref{fig:Tc1dist_c_2_3} we compare the distributions of $T_{3\rightarrow1}$ and $T_{2\rightarrow1}$ with different values of $p$.
Note that when $p=0.35$(Fig.~\ref{fig:Tc1dist_c_2_3}(a)), there is a sharp peak in the distribution of $T_{3\rightarrow1}$, which is due to finite size effect causing considerable amount of probability mass to enter region $\mathbb{B}$ (which is the union of region 1a,1b,1c shown in Fig.~\ref{fig:trans_prob_regions_p_0.35}).
Such peak does not exist for $T_{2\rightarrow1}$.
The distribution of $T_{3\rightarrow1}$ is a superposition of two distributions: a sharp peak due to attractors A, B, C and an exponential decay resulted from population relaxed from meta-stable states at point D, E, F (Refer to Fig.~\ref{fig:trans_prob_regions_p_0.35}).
From numerical integration of the master equation, we observe the effect of population size on $P_m(a)$, shown in Fig.~\ref{fig:Pm_p_0.370}. 
Most of the probability mass either concentrates near $a=N/2$ or at $a=1$.
As $N$ increases, the probability masses around $a=N/2$ and $a=1$ resemble two Delta functions, which we will exploit to make important approximations.
As $N\rightarrow\infty$, $P_{m}(a)$ can be approximated as
\begin{equation}
P_{m}(a)\approx\begin{cases}
b^\prime(N,p) & a=1\\
1-b^\prime(N,p) & a=N/2\\
0 & \text{otherwise}
\end{cases},\label{eq:P_m_approx}
\end{equation}
where $b^\prime(N,p)$ is the probability that the final state of process \rom{1} is $m=a=1$. 
This approximation holds when $p<0.5$.
$b^\prime(N,p)$ is shown in Fig.~\ref{fig:b_N_p}.
To simplify the notations below, we write $b^\prime(N,p)$ as $b^\prime$.

\begin{center}
\begin{figure}
\begin{centering}
\subfloat[\label{fig:Pm_N_300_p_0.370}$N=300$]{\begin{centering}
\includegraphics[width=0.47\columnwidth]{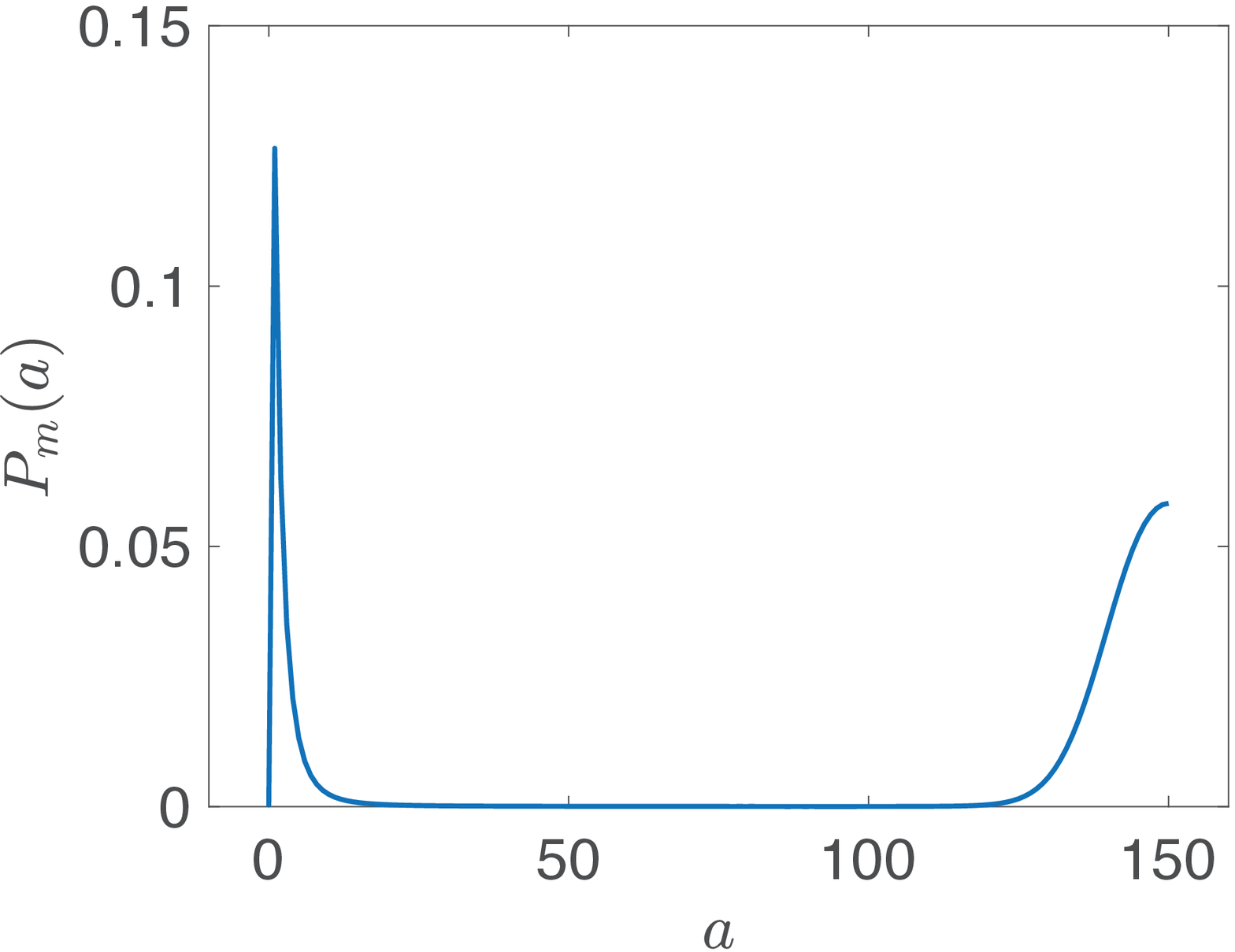}
\par\end{centering}

}\subfloat[\label{fig:Pm_N_900_p_0.370}$N=900$]{\begin{centering}
\includegraphics[width=0.47\columnwidth]{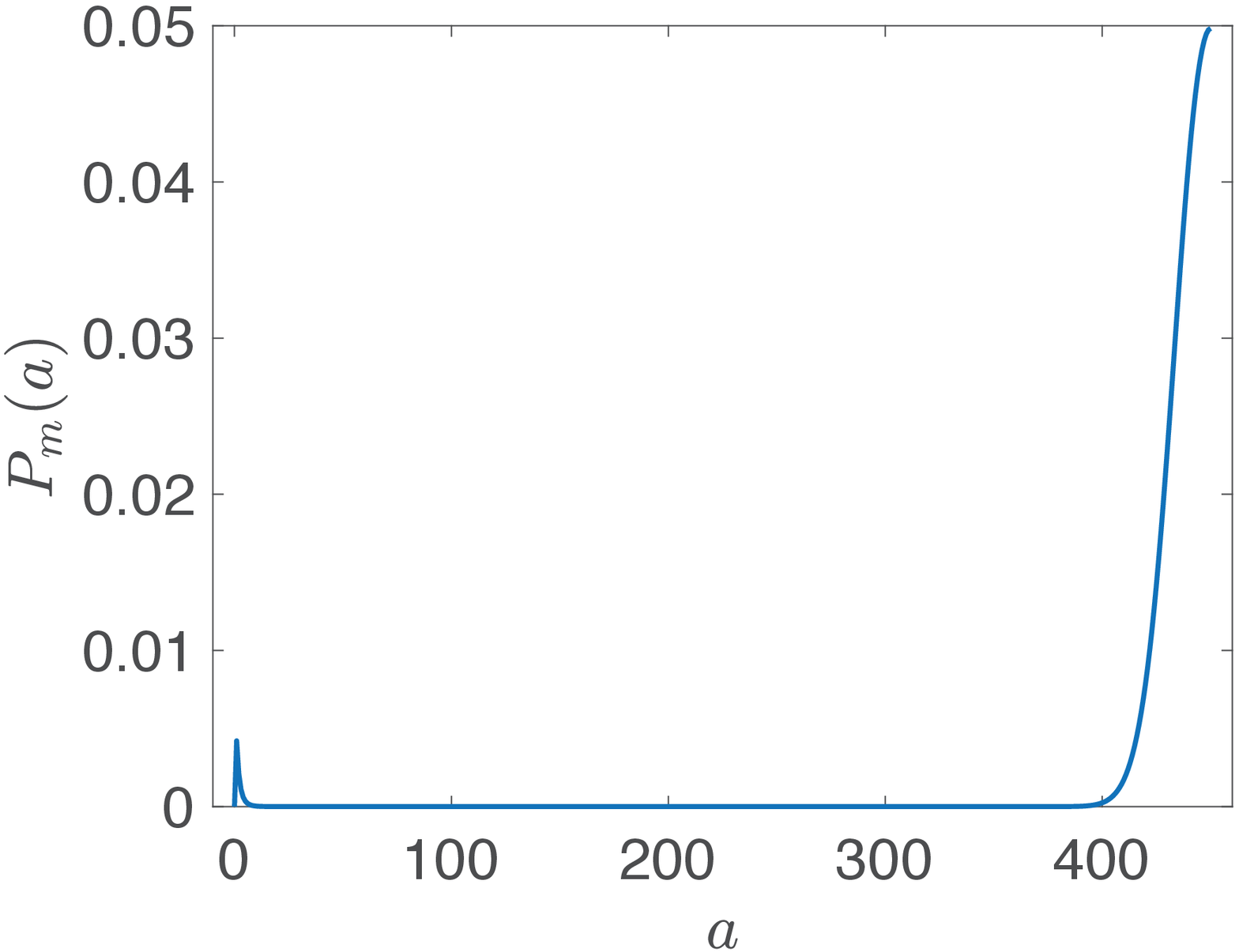}
\par\end{centering}

}
\par\end{centering}

\protect\caption{\label{fig:Pm_p_0.370}The effect of population size on $P_{m}(a)$. $p=0.37$.}

\end{figure}

\par\end{center}

With this approximation, 
\begin{equation}
\begin{aligned}
        &P_{3\rightarrow1}(T)\\
\approx & \int_0^Nda\int_0^T P_{3\rightarrow2,a}(T-t)P_{2\rightarrow1|a}(t)P_{m}(a)dt\\
\approx & \int_0^T P_{3\rightarrow2,1}(T-t)P_{m}(1)P_{2\rightarrow1|1}(t)dt+\\
        & \int_0^T P_{3\rightarrow2,\frac{N}{2}}\left(T-t\right)P_{m}\left(\frac{N}{2}\right)P_{2\rightarrow1|\frac{N}{2}}\left(t\right)dt\\
\approx & b^\prime P_{3\rightarrow2,1}(T)+\left[1-b^\prime\right]P_{2\rightarrow1|N/2}\left(T-\left< T_{3\rightarrow 2} \right>\right),
\end{aligned}
\end{equation}
where we make use of the numerical results that the distribution of $T_{3\rightarrow2}$ is narrowly peaked at $\left< T_{3\rightarrow 2} \right>$ and that $P_{2\rightarrow1|1}(t)\approx\delta(t-1)$, since for two states with one of the two opinions having $N-1$ votes and the other opinion having only one vote, convergence requires only one step.
Now we gain insights of the finite size effects of three-opinion system: there is a finite probability $b^\prime(N,p)$ that the population will reach the consensus state with a time scale that is independent of the population size; otherwise the intitial state of process \rom{2} is a polarized state, with a relaxation time scale that is an increasing function of the population size.

The average time to consensus $\left\langle T_{3\rightarrow1}\right\rangle $ is:
\begin{equation}
\begin{aligned}
\label{eq:avg_tttc}
  &\left<T_{3\rightarrow1}\right>\\
= & b^\prime \int_{0}^{\infty}t P_{3\rightarrow2,1}(t)dt+
   \left[1-b^\prime\right]\int_{\left< T_{3\rightarrow 2} \right>}^{\infty}t\, P_{2\rightarrow1|N/2}\left(t-\left< T_{3\rightarrow 2} \right>\right)dt \\
\approx & \left< T_{3\rightarrow 2} \right>+\left[1-b^\prime\right]\int_{0}^{\infty}t P_{2\rightarrow1|N/2}(t)dt,
\end{aligned}
\end{equation}
where we make the approximation that $\left< T_{3\rightarrow2,1}\right>\approx\left<T_{3\rightarrow2}\right>$ for the first term, and for the second term we change the limit of integral so that the $b^\prime$ factor in the first term combines with the second term to yield $\left<T_{3\rightarrow2}\right>$.

\begin{center}
\begin{figure}
\begin{centering}
\includegraphics[width=0.9\columnwidth]{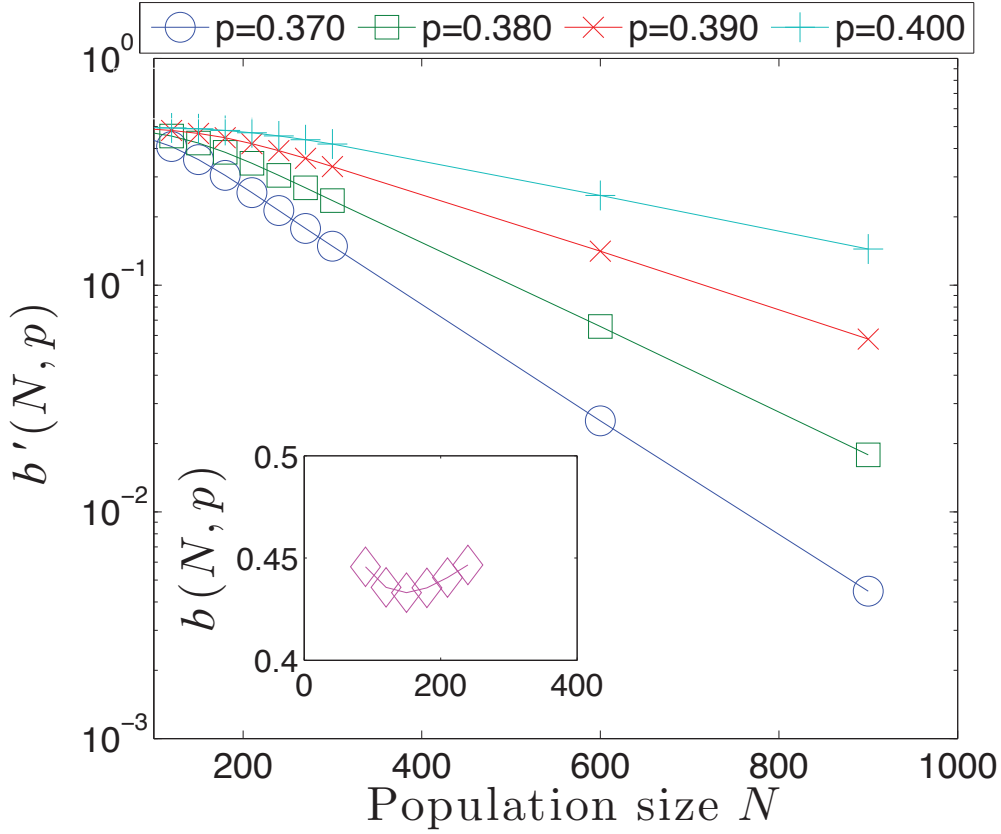}
\par\end{centering}

\protect\caption{\label{fig:b_N_p}
The probability that the final state of process I is $m=a=1$, is denoted here as $b^\prime(N,p)$ for $p<0.5$ and as $b(N,p)$ for $p>0.5$. The insert shows $b(N,p)$ when $p=0.65$.}

\end{figure}

\par\end{center}

According to Ref.~\cite{benczik_opinion_2009}, $\int_{0}^{\infty} t P_{2\rightarrow1|N/2}(t) dt \sim e^{a(p,M_{0})N}$.
Therefore, as $N\rightarrow\infty$, $\left<T_{3\rightarrow1}\right>$ is dominated by $e^{a(p,M_{0})N}$, where $a(p,M_{0})$ is positive, and the acceleration of the time to consensus will vanish. 
The dependence of $b^\prime(N,p)$ on $N$ and $p$ is shown in Fig.~\ref{fig:b_N_p}.
When $p<0.5$, $b^\prime(N,p)$ decreases exponentially with population size $N$. 
The difference in time to consensus between three state and two state model is small, due to the factor $b^\prime$ and the dominance of the second term in Eq.~\ref{eq:avg_tttc} for $p<0.5$.
We show in Fig.~\ref{fig:T2_T3_prediction_p0.370} the average time to consensus $\left< T_{3\rightarrow1}\right>$ for two-state model and three-state model when $p=0.37$, along with the predicted time to consensus for three-state model calculated using Eq.~\ref{eq:avg_tttc}.
Note that the prediction fits $\left\langle T_{3\rightarrow1}\right\rangle $ very well.

\begin{center}
\begin{figure}
\begin{centering}
\includegraphics[width=0.9\columnwidth]{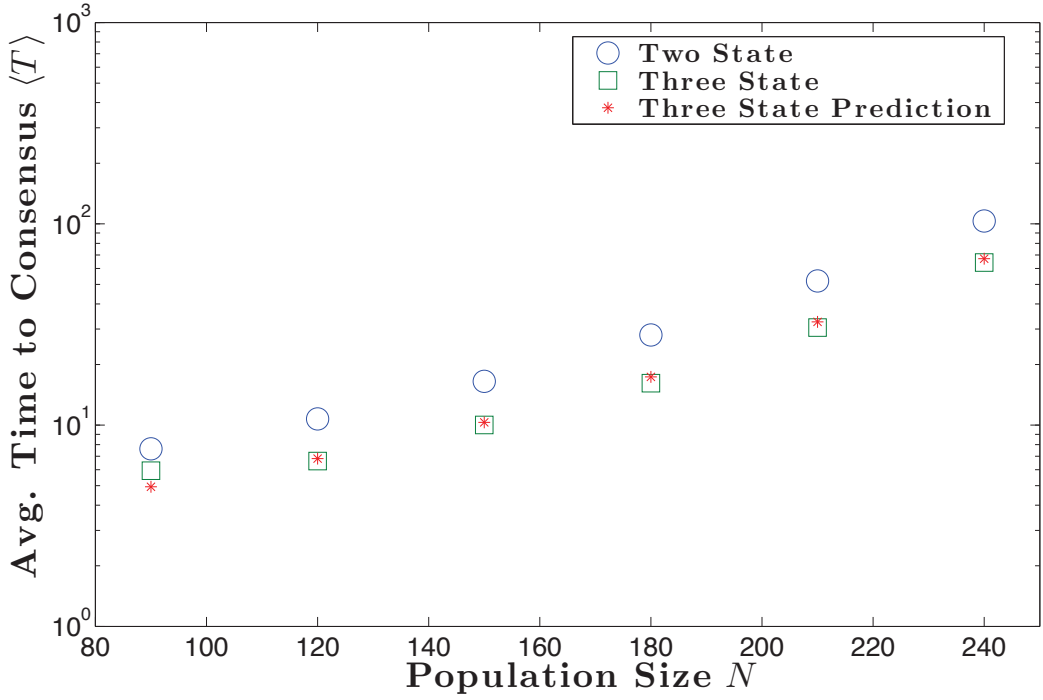}
\par\end{centering}

\protect\caption{\label{fig:T2_T3_prediction_p0.370}Average time to consensus for
two-state model, three-state model and that predicted by Eq.~\ref{eq:avg_tttc}.
$p=0.37$.}

\end{figure}

\par\end{center}

The case with $p>0.5$ should be treated differently.
(refer to the insert in Fig.~\ref{fig:b_N_p} for p>0.5) Numerical results show that the following is a reasonable approximation:
\begin{equation}
P_{m}(a)\approx\begin{cases}
b(N,p) & a=1\\
f(a) & N/2-\epsilon \le a \le N/2\\
0 & \text{otherwise}
\end{cases},\label{eq:P_m_approx_2}
\end{equation}
where $b(N,p)$ is the probability that the final state of process \rom{1} is $m=a=1$, and $\int_{N/2-\epsilon}^{N/2} f(a)da=1-b(N,p)$.
Therefore,
\begin{equation}
\begin{aligned}
	&P_{3\rightarrow1}(T)\\
	\approx & bP_{3\rightarrow2,1}(T)\\
 	& +\int_{N/2-\epsilon}^{N/2} \int^T_0 P_{3\rightarrow2,a}\left(T-t\right)f(a)P_{2\rightarrow1|a}\left(t\right)dt\, da\\
	\approx & bP_{3\rightarrow2,1}(T)+\int_{N/2-\epsilon}^{N/2} f(a)P_{2\rightarrow1|a}\left(T-\left<{T}_{3\rightarrow2}\right>\right)da,
\end{aligned}
\end{equation}
where, relative to $P_{2\rightarrow1|a}$, $P_{3\rightarrow2,a}$ can be regarded as a delta function centered at $\left<{T}_{3\rightarrow2}\right>$.
Therefore,
\begin{equation}
\begin{aligned}
&\left<T_{3\rightarrow1}\right>\\
= & b\left<{T}_{3\rightarrow2,1}\right>+ \int_{N/2-\epsilon}^{N/2} \int_{\left<{T}_{3\rightarrow2}\right>}^{\infty}f(a)t\, P_{2\rightarrow1|a}\left(T-\left<{T}_{3\rightarrow2}\right>\right)dt\,da \\
\approx & b\left<{T}_{3\rightarrow2}\right>+\int_{N/2-\epsilon}^{N/2}f(a)\left[\int_{0}^{\infty}t P_{2\rightarrow1|a}(t)dt+\left<{T}_{3\rightarrow2}\right>\right]da\\
= & \left<{T}_{3\rightarrow2}\right>+\int_{N/2-\epsilon}^{N/2} f(a)\int_{0}^{\infty}t P_{2\rightarrow1|a}(t)dt\, da\\
= & \left<{T}_{3\rightarrow2}\right>+\int_{N/2-\epsilon}^{N/2} f(a)\left\langle T_{2\rightarrow1|a}\right\rangle \, da\\
\le & \left<{T}_{3\rightarrow2}\right>+\int_{N/2-\epsilon}^{N/2} f(a)\left\langle T_{2\rightarrow1|N/2}\right\rangle \, da\\
= & \left<{T}_{3\rightarrow2}\right>+\left[1-b\right]\left\langle T_{2\rightarrow1|N/2}\right\rangle ,
\end{aligned}
\label{eq:avg_T_31_p_ge_05}
\end{equation}
where $\left\langle T_{2\rightarrow1|a}\right\rangle$ is effectively the time to consensus for two-state model given the initial condition is $m\equiv N_{\alpha}=a$, and the inequality follows from the fact that $\left\langle T_{2\rightarrow1|a}\right\rangle\le\left\langle T_{2\rightarrow1|N/2}\right\rangle$.
Therefore, Eq.~\ref{eq:avg_T_31_p_ge_05} gives an upper bound for the time to consensus when $p>0.5$. 
See Fig.~\ref{fig:T2_T3_prediction_p0.650} for $\left\langle T_{3\rightarrow1}\right\rangle $ when $p=0.65$. 
The value predicted by Eq.~\ref{eq:avg_T_31_p_ge_05} consistently serves as the upper bound for $\left\langle T_{3\rightarrow1}\right\rangle $. 
When $p=0.65$ , $\left<T_{3\rightarrow2}\right>\sim\exp(0.026N)$
while $\left<T_{2\rightarrow1}\right>\sim\exp(0.04N)$. The acceleration when
$p>0.5$ is a combination of two acceleration effects: 1) with probability
$b(N,p)$, $\left<T_{3\rightarrow1}\right>$ is dominated by $\left<T_{3\rightarrow2}\right>$,
which although depends on the population size exponentially, the
exponential coefficient is significantly smaller than that of $\left<T_{2\rightarrow1}\right>$.
2) with probability $1-b(N,p)$, the population reaches the consensus
state in a time scale at most $\left<T_{3\rightarrow2}\right>+\left\langle T_{2\rightarrow1}\right\rangle \approx\left\langle T_{2\rightarrow1}\right\rangle $
in large $N$ limit.

\begin{center}
\begin{figure}
\begin{centering}
\includegraphics[width=0.9\columnwidth]{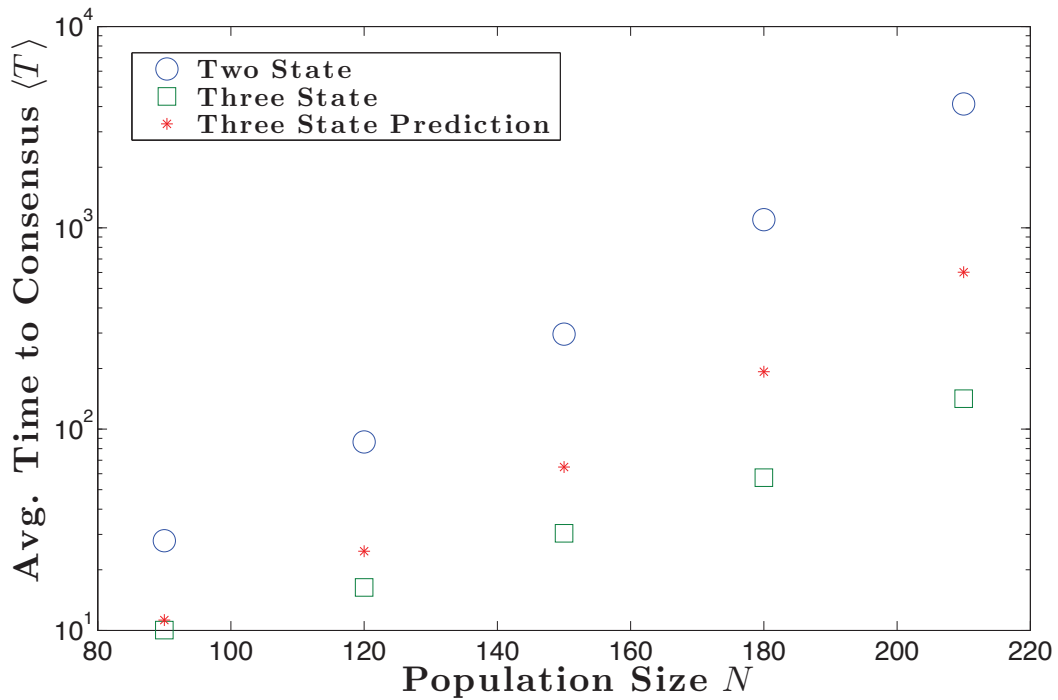}
\par\end{centering}

\protect\caption{\label{fig:T2_T3_prediction_p0.650}Average time to consensus for
two-state model, three-state model and that predicted by Eq.~\ref{eq:avg_tttc}.
$p=0.65$. $\left\langle T_{2\rightarrow1}\right\rangle \sim\exp(0.04N)$.}

\end{figure}

\par\end{center}
Note that $T_{2\rightarrow1}$ is an exponential decay with the approximately the same exponent, $\approx \exp(0.04N)$.
When $p=0.5$, the distributions are almost the same as expected (not shown).
Finite size effect also speed up the consensus reaching process when $p=0.65$.
\begin{figure}

	\subfloat[]{%
	\includegraphics[width=0.9\columnwidth]{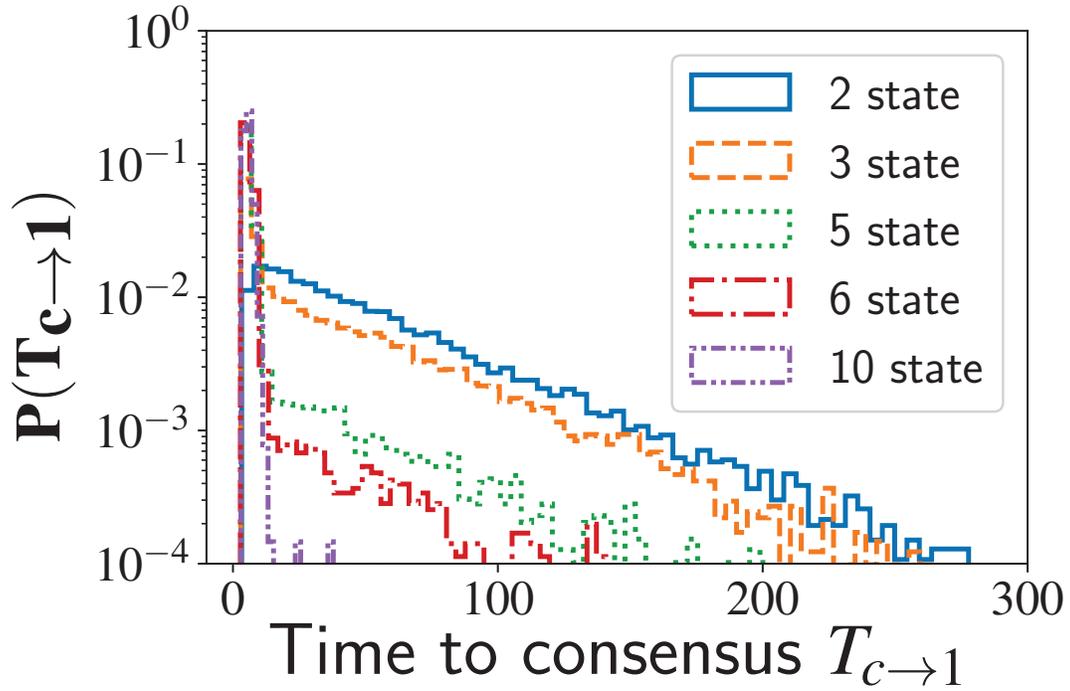}
	}\\
	\subfloat[]{%
	\includegraphics[width=0.9\columnwidth]{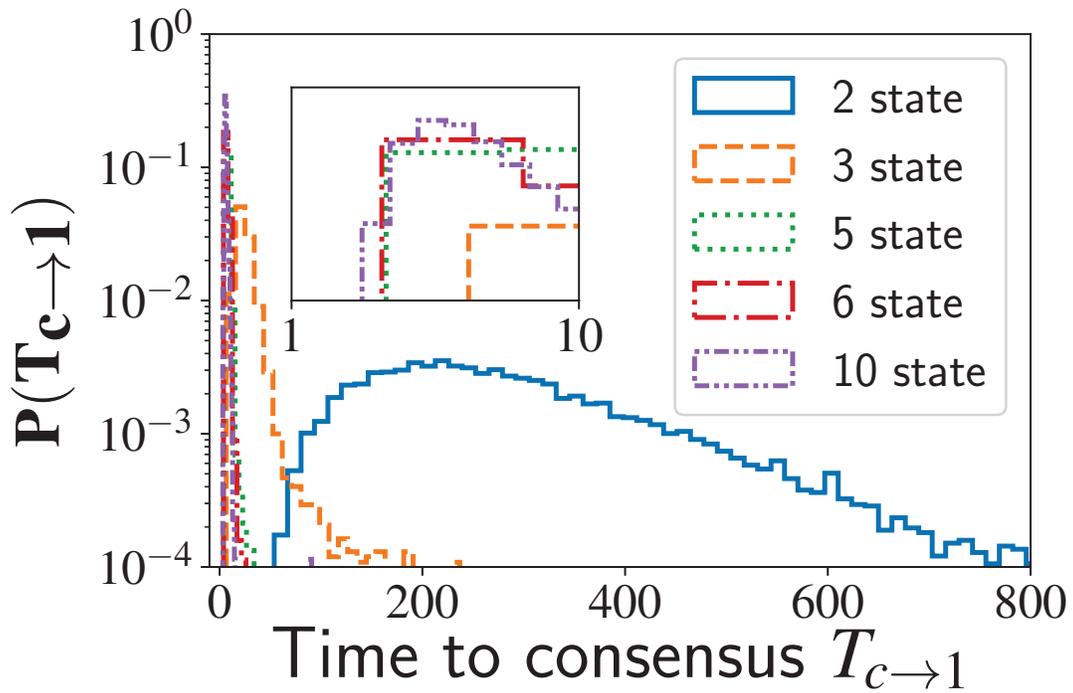}
	}

\protect\caption{\label{fig:Tc1dist_c_2_3_5_6_10}Distribution of time to consensus for two,three,five,six,ten-state system, obtained from Monte Carlo simulation. 
(a) $p=0.35$.
(b) $p=0.65$.
$N=150$.}
\end{figure}
We show in Fig.~\ref{fig:Tc1dist_c_2_3_5_6_10} the distribution of time to consensus for two,three,five,six,ten-state system, obtained from Monte Carlo simulation.
Our numerical results (not shown) show that for some values of $p$, the larger the number of available opinions is, the faster the time to convergence.

\section{Conclusions}

In summary, in thermodynamic limit, when $p<0.5$, a three-opinion population with equal number of voters in each opinion initially, which we defined as a totally fragmented initial condition, will spontaneously go to a two-state totally polarized state.
The difference in time to consensus between the two state and three state model is small in the large $N$ limit for $p<0.5$. (Refer to Fig.6a and Fig.11a)
When $p>0.5$, a totally fragmented three-opinion population will stay fragmented.
When intervention such as deliberate opinion conversion is allowed, the effort needed to push the fragmented three-opinion population to the consensus state, measured in minimal intervention cost, is less than that to push a polarized two-opinion population to the consensus state, when $p<0.8$.
In finite size system, by virtue of finite size effect, the totally fragmented three-opinion population will spontaneously reach the consensus state, with faster time to consensus, compared to polarized two-opinion population, for a broad range of $p$.
This rather counter-intuitive result has been analyzed using master equation and confirmed numerically.
Our analysis for opinion formation suggests an interesting application in social network beyond election, where time to consensus is critical when there is an imposed deadline. For example, in a multi-agent system, a company which markets the same product with three different brands may give a faster convergence of consumer behavior to one of the three brands.

\appendix

\section{\uppercase{Approximating} $W_{i\rightarrow j}$ \uppercase{in thermodynamics limit}}

Here we show how to approximate Eq.~\ref{eq:Wi2j} when $N\rightarrow\infty$.
$B(n,p)$ (Eq.~\ref{eq:BNp}) can be approximated by Gaussian distribution $\mathcal{N}(np,np(1-p))$ as $n\rightarrow\infty$ with $p$ held fixed. 

First, consider $w_{1\rightarrow2}\equiv \frac{N}{N_1}W_{1\rightarrow2}$. 
Rearrange the terms such that
\begin{equation}
\begin{aligned}
&w_{1\rightarrow2}(N_1,N_2)= \sum_{l'=0}^{N_2}B_{N_2,q}(l')\cdot\\
& \left[\sum_{l=0}^{N_1-1}\sum_{l''=0}^{N_3}B_{N_1-1,p}(l)B_{N_3,q}(l'')
 \Theta(l'-l)\Theta(l'-l'')\right],
\end{aligned}
\end{equation}
where $\Theta(\cdot)$ is unit step function.
The double sum in the square bracket can be approximated by the following integral:
\begin{equation}
\begin{aligned}
&I_{1\rightarrow2}(N_1,N_3;n_2)=\int_0^{N_1-1}\int_0^{N_3}\dfrac{1}{\sqrt{2\pi}\sigma_1}e^{-\frac{(n_1-(N_1-1)p)^2}{2\sigma_1^2}}\times\\
&\dfrac{1}{\sqrt{2\pi}\sigma_3}e^{-\frac{(n_3-zq)^2}{2\sigma_3^2}}\Theta(n_2-n_1)\Theta(n_2-n_3)dn_3dn_1,
\end{aligned}
\end{equation}
where $\sigma_1=\sqrt{(N_1-1)p(1-p)}$ and $\sigma_3=\sqrt{N_3q(1-q)}$.
With a change of variables $s\equiv n_1/N$, $s'\equiv n_2/N$ and $s''\equiv n_3/N$, the integral $I_{1\rightarrow2}$ becomes
\begin{equation}
\begin{aligned}
&I_{1\rightarrow2}(x,z;s')=\int_0^{x-\epsilon}\int_0^{z}\dfrac{1}{\sqrt{2\pi}\sigma_x}e^{-\frac{(s-(x-\epsilon)p)^2}{2\sigma_x^2}}\times\\
&\dfrac{1}{\sqrt{2\pi}\sigma_z}e^{-\frac{\left(s''-zq\right)^2}{2\sigma_3^2}}\Theta(s'-s)\Theta(s'-s'')ds''ds,
\end{aligned}
\end{equation}
where $\epsilon=1/N$, $\sigma_x=\sigma_1/N=\sqrt{\frac{(x-\epsilon)p(1-p)}{N}}$ and $\sigma_z=\sigma_3/N=\sqrt{\frac{zq(1-q)}{N}}$.
The integrand of the integral above is a product of two Gaussian distributions, with means $(x-\epsilon)p$ and $zq$, and variances $\frac{(x-\epsilon)p(1-p)}{N}$ and $\frac{zq(1-q)}{N}$, respectively.
Consider the first Gaussian distribution in $I_{1\rightarrow2}(x,z;s')$ as an example, in the limit where $N\rightarrow\infty$ with $x$ fixed, the mean $(x-\epsilon)p$ is constant up to an additive term of order $1/N$, and the variance vanishes in $1/N$ fashion.
Therefore, in the thermodynamic limit,
\begin{equation}
\begin{aligned}
&I_{1\rightarrow2}(x,z;s')=\int_0^{x-\epsilon}\int_0^{z}\delta(s-(x-\epsilon)p)\times\\
&\delta(s''-zq)\Theta(s'-s)\Theta(s'-s'')ds''ds,
\end{aligned}
\end{equation}
where $\delta(\cdot)$ is the Dirac delta function.
At this point, it is clear that if $s'<(x-\epsilon)p$ or $s'<zq$, $I_{1\rightarrow2}(x,z;s')=0$.
Therefore, $w_{1\rightarrow2}$ can be approximated by
\begin{equation}
w_{1\rightarrow2}(x,y,z)\approx\int_0^y I(x,z;s')\delta(s'-yq)ds'.
\end{equation}
Therefore, $w_{1\rightarrow2}$ is non-zero only when $y>(x-\epsilon)p$ and $y>zq$ simultaneously. 

Therefore, in the thermodynamics limit, $w_{1\rightarrow2}=1$ when $y>\max(p(x-\epsilon)/q,(1-x)/2)$ and zero otherwise. 
In other words, $W_{1\rightarrow2}=x/N$ when $y>\max(p(x-\epsilon)/q,(1-x)/2)$ and zero otherwise. 
The line $y=\max(p(x-\epsilon)/q,(1-x)/2)$, therefore, partitions the configuration space $(N_1,N_2)$ into \emph{positive region}, where $W_{1\rightarrow2}>0$ and \emph{zero region}, where $W_{1\rightarrow2}=0$.
Approximation of other transition probabilities can be made in similar ways.

\section{\uppercase{Partitioning the configuration space based on } $W_{i\rightarrow j}$}

Although the right hand side master equation, or Eq.~\ref{eq:master_eq_short}, involves six transition probabilities $W_{i\rightarrow j}$, at many points in the configuration space $(N_1,N_2)$, as we have demonstrated in the appendix, transition probabilities are either zero or have linear dependence in only one of the three variables, $x$, $y$ or $z$, in the thermodynamic limit.
Therefore, the qualitative behavior of the master equation in the neighborhood of a given point $(n_1,n_2)$ is dictated by transition probabilities that are positive in the neighborhood of $(n_1,n_2)$.
For this reason, it is useful to partition the configuration space according to what the set of positive transition probabilities is.

To facilitate the formal investigation, we introduce the following notations.
First we denote the positive region of $W_{i\rightarrow j}$ by 
\begin{equation}
\begin{aligned}
\mathbb{F}_{i\rightarrow j}=\{(x,y)|(x,y)\in\Omega,W_{i\rightarrow j}(x,y)>0\}
\end{aligned}
\end{equation}
where $\Omega\equiv\{(x,y)|0\le x\le 1,0\le y\le 1,x+y\le 1\}$ is essentially the configuration space.
The zero region of $W_{i\rightarrow j}$ is therefore 
\begin{equation}
\begin{aligned}
\mathbb{F}^{C}_{i\rightarrow j}\equiv\Omega\setminus\mathbb{F}_{i\rightarrow j}.
\end{aligned}
\end{equation}
Let $\Gamma=\{(1\rightarrow2),(1\rightarrow3),(2\rightarrow1),(2\rightarrow3),(3\rightarrow1),(3\rightarrow2)\}$, which is the set of labels of all transition probabilities in three-state system.
We use $\gamma\in 2^{\Gamma}$ to specify the set of positive transition probabilities.
The subset of configuration space where only $\{W_{i\rightarrow j}|(i\rightarrow j)\in\gamma\}$ are positive is denoted by
\begin{equation}
\mathbb{A}(\gamma)=
	\left[\bigcap_{(i\rightarrow j)\in\gamma}\mathbb{F}_{i\rightarrow j}\right]
		\bigcap
	\left[\bigcap_{(i\rightarrow j)\in\Gamma\setminus\gamma}\mathbb{F}^{C}_{i\rightarrow j}\right],
\end{equation}
which we will call \emph{asymptotic region of $\gamma$}.
\begin{table}[h]
\centering
\begin{tabular}{c | c}
	Region Name in Fig.~\ref{fig:trans_prob_regions_p_0.35} & Corresponding Asymptotic Region \\
	\hline
	1a & $\mathbb{A}\left[1\rightarrow2,3\rightarrow2\right]$ \\
	1b & $\mathbb{A}\left[2\rightarrow1,3\rightarrow1\right]$ \\
	1c & $\mathbb{A}\left[2\rightarrow3,1\rightarrow3\right]$ \\
	2a & $\mathbb{A}\left[3\rightarrow2,2\rightarrow3,1\rightarrow2\right]$ \\
	2b & $\mathbb{A}\left[2\rightarrow1,3\rightarrow2,1\rightarrow2\right]$ \\
	2c & $\mathbb{A}\left[3\rightarrow1,2\rightarrow1,1\rightarrow2\right]$ \\
	2d & $\mathbb{A}\left[1\rightarrow3,2\rightarrow1,3\rightarrow1\right]$ \\
	2e & $\mathbb{A}\left[3\rightarrow1,1\rightarrow3,2\rightarrow3\right]$ \\
	2f & $\mathbb{A}\left[3\rightarrow2,1\rightarrow3,2\rightarrow3\right]$ \\
\end{tabular}
\caption{Region names in Fig.~\ref{fig:trans_prob_regions_p_0.35} and their corresponding asymptotic region.}\label{tab:trans_prob_regions_p_0.35}
\end{table}
It can be shown that $\{\mathbb{A}(\gamma)|\gamma\in 2^{\Gamma}\}$ is a partition of the configuration space.
Fig.~\ref{fig:trans_prob_regions_p_0.35} shows the partition according to $\mathbb{A}(\gamma)$ when $p=0.35$.
By definition, there are 64 distinct asymptotic regions, but for a particular value of $p$, many regions are empty sets.
Some asymptotic regions are so small they do not play much a role in the behaviors of the master equation.
Table~\ref{tab:trans_prob_regions_p_0.35} lists the asymptotic regions that are prominent when $p=0.35$.
See Fig.~\ref{fig:trans_prob_regions_p_0.65} for $W_{1\rightarrow2}$ when $p=0.65$ (refer to Table \ref{tab:trans_prob_regions_p_0.65}). 
Compared to the case when $p=0.35$, regions 2a to 2f that are prominent in Fig.~\ref{fig:trans_prob_regions_p_0.35} vanish.
On the other hand, regions 3a to 3f, which are not visible in Fig.~\ref{fig:trans_prob_regions_p_0.35}, become prominent.
The white area in the middle of the configuration space is $\mathbb{A}(\emptyset)$.
\begin{table}[h]
\centering
\begin{tabular}{c | c}
	Region Name in Fig.~\ref{fig:trans_prob_regions_p_0.65} & Corresponding Asymptotic Region \\
	\hline
	3a & $\mathbb{A}(1\rightarrow2)$\\
	3b & $\mathbb{A}(1\rightarrow3)$\\
	3c & $\mathbb{A}(2\rightarrow3)$\\
	3d & $\mathbb{A}(2\rightarrow1)$\\
	3e & $\mathbb{A}(3\rightarrow1)$\\
	3f & $\mathbb{A}(3\rightarrow2)$\\
	white area & $\mathbb{A}(\emptyset)$\\
\end{tabular}
\caption{Regions that are prominent in Fig.~\ref{fig:trans_prob_regions_p_0.65} but not prominent in Fig.~\ref{fig:trans_prob_regions_p_0.35}.\label{tab:trans_prob_regions_p_0.65}
Their corresponding asymptotic regions are indicated.}
\end{table}

\section{\uppercase{Deriving master equation in regions }1a, 1b, 1c\uppercase{ of the configuration space}}

Since region 1a of Fig.~\ref{fig:trans_prob_regions_p_0.35} is $\mathbb{A}\left[1\rightarrow2,3\rightarrow2\right]$, $W_{1\rightarrow2}$ and $W_{3\rightarrow2}$ are positive and proportional to $x$ and $z$, respectively, while other transition probabilities are zero.
Therefore, the master equation can be approximated by:
\begin{equation}
\begin{aligned}
&\partial_{t}P(x,y)\\
= & -\left[x+z\right]P(x,y)+(x+\epsilon)P(x+\epsilon,y-\epsilon)\\
  & +(z+\epsilon)P(x,y-\epsilon)\\
= & -\left[x+z\right]P(x,y)\\
  & +(x+\epsilon)\left[P(x,y)+\frac{\partial P}{\partial x}\epsilon+\frac{\partial P}{\partial y}(-\epsilon)+\cdots\right]\\
  & +(z+\epsilon)\left[P(x,y)+\frac{\partial P}{\partial y}(-\epsilon)+\cdots\right]\\
\approx & \epsilon\left[x\frac{\partial P}{\partial x}+(y-1)\frac{\partial P}{\partial y}+2P(x,y)\right],
\end{aligned}
\end{equation}
where $\epsilon=1/N$ and the last approximation keeps only the 1st order terms of $\epsilon$.

By the method of characteristics, one can obtain $x(t)=x_0\exp(-t)$ and $y(t)=(y_0-1)\exp(-t)+1$. 
The trajectory of a delta probability mass centered at $(x_0,y_0)$ is therefore $y=(y_0-1)\frac{x}{x_0}+1$.
It follows that $z(t)=1-x(t)-y(t)=z_0\exp(-t)$.
By symmetry, one can read out $x(t),y(t),z(t)$ in regions 1b and 1c.

\section{\uppercase{Deriving master equation in regions }2a, 2b, 2c, 2d, 2e, 2f\uppercase{ of the configuration space when }$p<0.5$}

Similarly, in region~2a, or $\mathbb{A}\left[1\rightarrow2,2\rightarrow3,3\rightarrow2\right]$, only $W_{1\rightarrow2},W_{2\rightarrow3}$ and $W_{3\rightarrow2}$ are positive, and the master equation in that region can be approximated by, in the $N\rightarrow\infty$ limit, 
\begin{align}
\begin{split}
&\partial_{t}P(x,y)\\
= & -\left[x+y+z\right]P(x,y)+(x+\epsilon)P(x+\epsilon,y-\epsilon)\\
&+(y+\epsilon)P(x,y+\epsilon)+(z+\epsilon)P(x,y-\epsilon)\\
\approx & \epsilon\left[3P(x,y)+x\frac{\partial P}{\partial x}+(2y-1)\frac{\partial P}{\partial y}\right].
\end{split}
\end{align}
Using the method of characteristics, one can find $y(t)=1/2-(1/2-y_0)e^{-2t}$ and $x(t)=x_0e^{-t}$. 
The trajectory of a delta probability mass centered at $(x_0,y_0)$ is therefore 
\begin{equation}
\label{eq:traj_2a}
y=\frac{1}{2}-\frac{1-2y_0}{2x_0^2}x^2.
\end{equation}
It follows that $z(t)=1-x(t)-y(t)=\frac{1}{2}-x_0\exp(-t)+(\frac{1}{2}-y_0)\exp(-2t)$ or $z(t)=z_0\exp(-2t)+x_0[\exp(-2t)-\exp(-t)]+\frac{1}{2}[1-\exp(-2t)]$.

Since region 2f is $\mathbb{A}[3\rightarrow2,1\rightarrow3,2\rightarrow3]$, the master equation can be approximated by
\begin{align}
\begin{split}
&\partial_{t}P(x,y)\\
\approx & \epsilon\left[3P(x,y)+x\frac{\partial P}{\partial x}+(2y+x-1)\frac{\partial P}{\partial y}\right].
\end{split}
\end{align}
By method of characteristics, one can get $x(t)=x_0\exp(-t)$ and $y(t)=y_0\exp(-2t)+x_0[\exp(-2t)-\exp(-t)]+\frac{1}{2}[1-\exp(-2t)]$.
It follows that $z(t)=1-x(t)-y(t)=\frac{1}{2}-(\frac{1}{2}-z_0)\exp(-2t)$.
Note the symmetry beween $y(t)$ in region 2a and $z(t)$ in region 2f.

By symmetry, one can read out $x(t),y(t),z(t)$ in regions 2b, 2c, 2d and 2e.

\section{\uppercase{Deriving master equation in regions }3a, 3b, 3c, 3d, 3e, 3f\uppercase{ of the configuration space when }$p>0.5$}

In region~3a, or $\mathbb{A}\left[1\rightarrow2\right]$, the master equation can be approximated by
\begin{equation}
\begin{aligned}
	&\partial_t P(x,y)\\
	\approx&\epsilon\left[P(x,y)-x\dfrac{\partial P}{\partial y}+x\dfrac{\partial P}{\partial x}\right].
\end{aligned}
\end{equation}
Using the method of characteristics, one will obtain $x=x_0\exp(-t)$, $y=y_0+x_0(1-\exp(-t))$ and $z=1-x_0-y_0$.

\section{\uppercase{Deriving Minimal Intervention Cost}}

Let us consider the case where there are $c$ opinions.
Suppose the initial opinion distribution is $\bold{X}_0\equiv(x_{1,0},x_{2,0},\cdots,x_{c,0})$ and the desired opinion distribution is $\bold{X}_1\equiv(x_{1,1},x_{2,1},\cdots,x_{c,0})$.
Conservation of the number of opinion holders leads to
\begin{equation}
	\sum_{j=1,j\neq i}^c \Delta_{j\rightarrow i} - \sum_{j=1,j\neq i}^c \Delta_{i\rightarrow j} = \Delta x_{i}\equiv x_{i,1}-x_{i,0},
\end{equation}
for $i=1,\cdots,c$.
The minimal intervention cost for $c$ opinions is thus
\begin{equation}
\begin{aligned}
	& & & \delta n_c(\bold{X}_0\rightarrow\bold{X}_1) = \\
	& \underset{\{\Delta_{ij}\}}{\text{min}}
	& & \sum_{i=1,j=1,i\neq j}^c\Delta_{i\rightarrow j} \\
	& \text{subject to}
	& & \sum_{j=1,j\neq i}^c \Delta_{j\rightarrow i} - \sum_{j=1,j\neq i}^c \Delta_{i\rightarrow j} = \Delta x_{i}, 
\end{aligned}
\end{equation}
for $i=1,\cdots,c$.

We can prove that
\begin{equation}
\begin{aligned}
	\delta n_c(\bold{X}_0\rightarrow\bold{X}_1) =\frac{1}{2}\|\bold{X}_1-\bold{X}_0\|_1=\frac{1}{2}\sum_{i=1}^{c}|x_{i,1}-x_{i,0}|.
\end{aligned}
\end{equation}
Consider the case where $\Delta x_i\neq0\;\forall i$.
Group the indexes of $\Delta x_i$ according to their signs: $\mathbb{I}^+=\{i|\Delta x_i>0\}$ and $\mathbb{I}^-=\{i|\Delta x_i<0\}$.
Since $\sum_{i=\mathbb{I}^+}\Delta x_i=-\sum_{i=\mathbb{I}^-}\Delta x_i$ (conservation of total number of opinions), $\sum_{i,j=1,i\neq j}^c\Delta_{i\rightarrow j}$ reaches minimum when we set $\sum_{j\in\mathbb{I}^+}\Delta_{i\rightarrow j}=-\Delta x_i\quad\forall i\in\mathbb{I}^-$, $\Delta_{i\rightarrow j}=0\quad\forall i,j\in\mathbb{I}^-,i\neq j$, and $\Delta_{i\rightarrow j}=0,\forall i\in\mathbb{I}^+$.
Therefore, 
\begin{equation}
\begin{aligned}
	&\sum_{i,j=1,i\neq j}^c\Delta_{i\rightarrow j}\\
	=&\sum_{i\in\mathbb{I}^-}\sum_{j\in\mathbb{I}^+}\Delta_{i\rightarrow j}=\sum_{i\in\mathbb{I}^-}-\Delta x_i=\sum_{i\in\mathbb{I}^+}\Delta x_i\\
	=&\frac{1}{2}\left(\sum_{i\in\mathbb{I}^-}-\Delta x_i+\sum_{i\in\mathbb{I}^+}\Delta x_i\right)\\
	=&\frac{1}{2}\left(\sum_{i\in\mathbb{I}^-}|\Delta x_i|+\sum_{i\in\mathbb{I}^+}|\Delta x_i|\right)\\
	=&\frac{1}{2}\left(\sum_{i=1}^c|\Delta x_i|\right)=\frac{1}{2}\|\bold{X}_1-\bold{X}_0\|_1
\end{aligned}
\end{equation}

\section{\uppercase{Calculating minimal intervention cost in various scenarios}}

If the spontaneous dynamics of metastable states' basins of attraction (2a-2f, 3a-3f) is not taken into account, the best way to force the consensus state is to move the opinion distribution from (1/3,1/3,1/3) to $(\frac{1-p}{1+p},\frac{p}{1+p},\frac{p}{1+p})$, $(\frac{p}{1+p},\frac{1-p}{1+p},\frac{p}{1+p})$ or $(\frac{p}{1+p},\frac{p}{1+p},\frac{1-p}{1+p})$, reaching the boundary of the basins of attraction of the consensus states, while circumventing the basins of attraction of the meta-stable states.
When $0<p<0.5$, the minimal intervention cost $\delta n_{3\rightarrow1}=\frac{2-4p}{3(p+1)}$, according to Eq.~{\ref{eq:intervention_cost}. 
When $0.5<p<1$, $\delta n_{3\rightarrow1}=\frac{4p-2}{3(p+1)}$.

When we exploit the spontaneous dynamics of metastable states' basins of attraction (2a-2f, 3a-3f), the resulted minimal intervention cost $\delta n_{3\rightarrow1}^h\le\delta n_{3\rightarrow1}$.
When $p<\sqrt{6}-2$, we can demonstrate the process in Fig.~\ref{fig:hitchhike_p_0.2}: the path that goes from (1/3,1/3,1/3) to one of the consensus state, represented by point B, can be broken into three stages.
In the first stage, the population evolves according to $x_{2f}(t)=\frac{1}{3}e^{-t},z_{2f}(t)=\frac{1}{2}−(\frac{1}{2}-\frac{1}{3})e^{-2t}$, which is represented by the red segment.
Once the population reaches the point $(x_{2f}(t^*),z_{2f}(t^*))$ such that
\begin{equation}
	t^*,x^*=\underset{t,x}{\operatorname{argmin}}\,\delta n_3(x_{2f}(t)\rightarrow x,z_{2f}(t)\rightarrow z_{2f|1c}(x)),
\end{equation}
where $y_{2f|1c}(x)$ is the line separating region 2f from 1c, one should intervene and send the population from $(x_{2f}(t^*),z_{2f}(t^*))$ to $(\frac{p}{1+p},\frac{p}{1+p},\frac{1-p}{1+p})$, and the minimal intervention cost is half of the $L_1$ norm of the blue segment.
Afterwards, the spontaneous dynamics will send the population to the consensus state marked by B.

When $\sqrt{6}-2<p<0.5$, the spontaneous dynamics sends the population right into the consensus state without any need for intervention, and hence $\delta n_{3\rightarrow1}^h=0$.

When $p>0.5$, since the population is frozen in the initial state, one should immediately intervene.
The intervention strategy is portrayed in Fig.~\ref{fig:hitchhike_p_ge_0.5}.
$\delta n^h_{3\rightarrow1}<\delta n_{2\rightarrow1}$ for $0.5<p\lesssim0.8$.

\bibliography{mylib}

\end{document}